\documentclass[sigconf]{acmart}

\usepackage{multirow}

\usepackage{microtype}
\usepackage{graphicx}
\usepackage{booktabs} 
\usepackage{hyperref}
\usepackage{subcaption}
\usepackage{mathtools}
\usepackage{setspace} 
\usepackage{caption}

\usepackage{balance}
\usepackage{adjustbox}
\usepackage{makecell}
\usepackage{amsfonts}

\usepackage[hang]{footmisc}

\newcommand{\parjump}{\vspace{+0.5em}} 

\newcommand{\secspace}{\vspace{-0.3em}}
\newcommand{\subsecspace}{\vspace{-0.2em}}

\newcommand{\system}{Mic2Mic}

\setcopyright{none}
\renewcommand\footnotetextcopyrightpermission[1]{}
\newcommand{\squishlist}{
 \begin{list}{$\bullet$}
  { \setlength{\itemsep}{0pt}
     \setlength{\parsep}{3pt}
     \setlength{\topsep}{3pt}
     \setlength{\partopsep}{0pt}
     \setlength{\leftmargin}{1em}
     \setlength{\labelwidth}{1em}
     \setlength{\labelsep}{0.5em} } }

\newcommand{\squishend}{
  \end{list}  }

\begin{document}

\title[Mic2Mic: CycleGANs for Robust Speech Systems]{Mic2Mic: Using Cycle-Consistent Generative Adversarial Networks to Overcome Microphone Variability in Speech Systems}

\vspace{-3cm}

\author{
{Akhil Mathur$^{\ddagger}$$^{\dagger}$, Anton Isopoussu$^{\ddagger}$, Fahim Kawsar$^{\ddagger}$, Nadia Berthouze$^{\dagger}$, Nicholas D. Lane$^{\diamond}$
}}
\affiliation{%
$^{\ddagger}$Nokia Bell Labs, $^{\dagger}$University College London, $^{\diamond}$University of Oxford
}

\renewcommand{\shortauthors}{Mathur et al.}

\begin{abstract}
\small{
\noindent
Mobile and embedded devices are increasingly using microphones and audio-based computational models to infer user context. A major challenge in building systems that combine audio models with commodity microphones is to guarantee their accuracy and robustness in the real-world. Besides many environmental dynamics, a primary factor that impacts the robustness of audio models is \emph{microphone variability}. In this work, we propose Mic2Mic -- a \emph{machine-learned system component} -- which resides in the inference pipeline of audio models and at real-time reduces the variability in audio data caused by microphone-specific factors. Two key considerations for the design of Mic2Mic were: a) to decouple the problem of microphone variability from the audio task, and b) put minimal burden on end-users to provide training data. With these in mind, we apply the principles of cycle-consistent generative adversarial networks (CycleGANs) to learn Mic2Mic using unlabeled and unpaired data collected from different microphones. Our experiments show that Mic2Mic can recover between 66\% to 89\% of the accuracy lost due to microphone variability for two common audio tasks.
}

\end{abstract}


\maketitle

{\secspace}
\section{Introduction}
\label{sec:intro}
{\secspace}
Recent advances in audio-based computational models have enabled a number of audio sensing applications on wearable and embedded devices. Past works have shown the feasibility of using audio signals to infer eating activities\cite{amft2005analysis, amft2006methods}, ambient conditions~\cite{xu2013crowd}, subjective user states \cite{anagnostopoulos2015features}, and productivity~\cite{lee2013sociophone, tan2013sound}.

A major challenge to widespread deployment and usability of audio models is to maintain high accuracy and robustness in real-world scenarios. Past works have looked at making audio models robust against background noise~\cite{xu2014experimental, qian2016very}, room reverberations~\cite{kim2016environmental}, and speaker identities~\cite{senior2014improving}. However, considerably less attention has been paid to make audio models more robust to the challenge of microphone variability. Das et al.\cite{das2014fingerprinting} revealed surprising findings that microphone variabilities across different smartphones are computationally so significant that it is even possible to fingerprint a smartphone based on its microphone. Our own experiments (in \S~\ref{subsec.emipiral}) also validate the presence of microphone variabilities in embedded microphones and show how they alone can degrade the accuracy of audio models by up to 15\%. 

A microphone's performance is characterized by a number of parameters such as its frequency response, impedance rating, output level, and even the signal processing applied on the raw audio before it is made available to user applications. All these parameters contribute to the \emph{transfer function} of a microphone, and affect how a physical audio signal is converted into digital output by the microphone. Interestingly, different microphones have different transfer functions owing to the variations in their underlying hardware and software processing pipelines -- as such, for the same physical audio signal, variations in the digital outputs of different microphones are likely. From the perspective of machine learning models, microphone variability can lead to \emph{domain shift} or mismatch between the training and test domains. As such, if an audio model is \emph{deployed} on a microphone whose properties (or transfer function) differ from the microphone(s) used for \emph{training} the model, this domain mismatch can lead to poor inference performance.

An easy solution to this problem is to control the variability in microphone hardware between training and deployment phases. This can be done by either restricting the audio application work only on a specific type of microphones, or by collecting training data from all types of microphones that will be encountered in the deployment phase. None of these approaches are ideal: firstly, many audio processing services are transitioning to an API-as-a-service model~\cite{alexaapi, googleapi} to allow for integration with any off-the-shelf microphone. Secondly, collecting audio training data from all possible types of microphones is an expensive and time-consuming process, and is certainly not feasible for most developers. 

In this paper, we present a practical solution to recover the accuracy of audio models otherwise lost due to microphone variability. We propose to frame the problem of microphone variability as an \emph{audio translation} problem, that is, given a set of audio data from a source microphone, can we \emph{translate} it such that it will resemble data collected from a target microphone? More formally, if $y$ and $y'$ are the recordings of the same audio signal from two microphones $A$ and $B$ respectively, we would like to learn a microphone translation function $f$, such that $y' = f (y)$. 

Our solution, \emph{Mic2Mic}, is a machine-learned system component which runs on embedded devices and at inference-time \emph{translates} data from the test microphone domain to the training microphone domain, thereby reducing the domain shift. More importantly, Mic2Mic decouples the microphone variability problem from the downstream audio task (e.g., ASR), and provides an inference-time solution which audio model developers can simply import in their inference pipeline to solve for microphone variability.  

A major practical hurdle in learning the translation function $f$ is the difficulty of obtaining paired or aligned audio data from multiple microphones -- in \S~\ref{sec:cycle} we discuss how Mic2Mic solves this challenge by applying the principles of cycle-consistent generative adversarial networks (CycleGAN) to learn a translation function using \emph{unlabeled} and \emph{unpaired} data. In \S~\ref{sec:inference}, we present the training and inference architecture of Mic2Mic and finally, in \S~\ref{sec:eval} we evaluate how Mic2Mic performs on two popular audio modeling tasks, namely Keyword Detection and Emotion Recognition. The main contributions of this work include:

\squishlist {
\item{A new perspective of looking at microphone variability as an \emph{audio data translation} problem, and developing a machine-learned system component called \system{} to improve the accuracy of audio models on unseen microphones.}

\item{A systematic study highlighting microphone variability on embedded devices using state-of-the-art audio models and a careful examination of the design space of \system{}.}

\item{The development of an audio translation model using unpaired and unlabeled data from multiple microphones by applying the principles of CycleGAN.}

\item{A comprehensive evaluation of \system{} on two audio tasks, namely Keyword Detection and Emotion Recognition. Mic2Mic outperforms all tested baselines in solving the problem of microphone variability.
}
}
\squishend

{\secspace}
\section{Background}
\label{sec:bg}
{\secspace}
\label{subsec.emipiral}
\noindent

Model robustness has been a prominent topic of research in the speech community~\cite{xu2014experimental, qian2016very, senior2014improving}. A fundamental challenge to robustness of audio models is associated with the variability in microphones used to record the audio signal. Prior works~\cite{das2014fingerprinting, ipsn18_mathur} have also alluded to this phenomenon in the context of smartphone and smartwatch microphones. 

In this section, we focus our attention on embedded microphones and provide intuition on how variabilities may get introduced in the audio data collected by them. Next, we present an experiment to validate the data variabilities across microphones, by controlling confounding factors such as variations in input signals and room acoustics. Finally, we evaluate the impact of these variabilities on a state-of-the-art audio model. We conclude by highlighting existing approaches to solving this problem, discussing their limitations, and motivating our proposed solution.

\begin{figure}[htb]
\centering
\includegraphics[width=\linewidth]{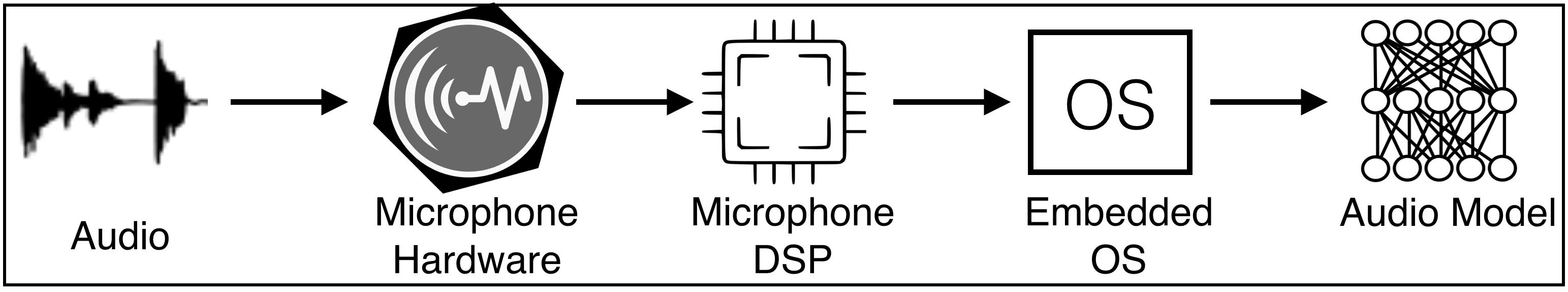}
\vspace{-0.5cm}%
\caption{Sensing and inference pipeline for audio models.}
\vspace{-0.4cm}
\label{diag:processing_pipeline}
\end{figure}

\parjump{}
\noindent
\textbf{Audio Processing Pipeline.} As shown in Figure~\ref{diag:processing_pipeline}, before an audio signal even reaches the audio classifier, it goes through a number of processing stages. Firstly, the physical audio signal is captured by the acoustic sensor of the microphone and converted into an electronic and then digital signal. Thereafter, the signal is processed by a Digital Signal Processor (DSP) on the embedded device where audio enhancement techniques such as noise filtering and delay-and-sum beamforming are applied. Finally, the processed signal is exposed to user applications such as a pre-trained audio classifier to compute task-specific inferences.

\begin{figure*}[htb]
\centering
\begin{subfigure}[b]{0.8\linewidth}
\includegraphics[width=\linewidth]{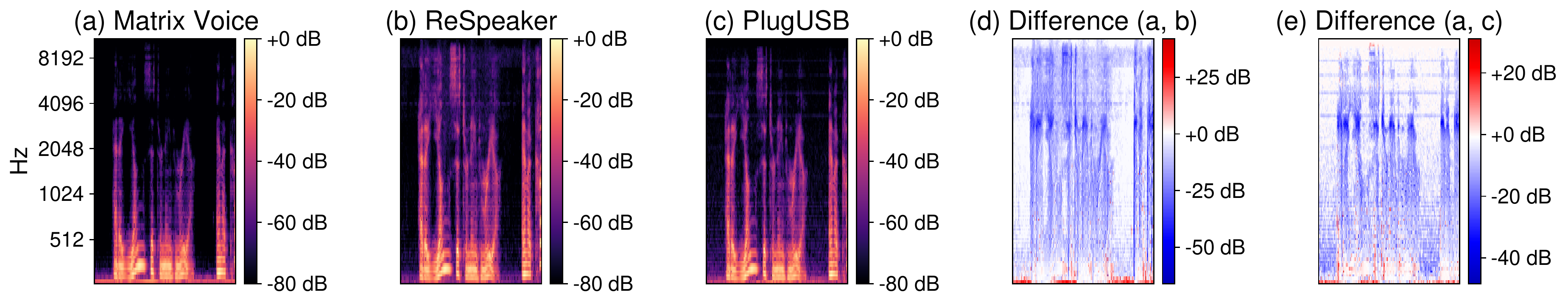}
\vspace{-0.4cm}
\caption{}
\end{subfigure}
\begin{subfigure}[b]{0.15\linewidth}
\includegraphics[width=\linewidth]{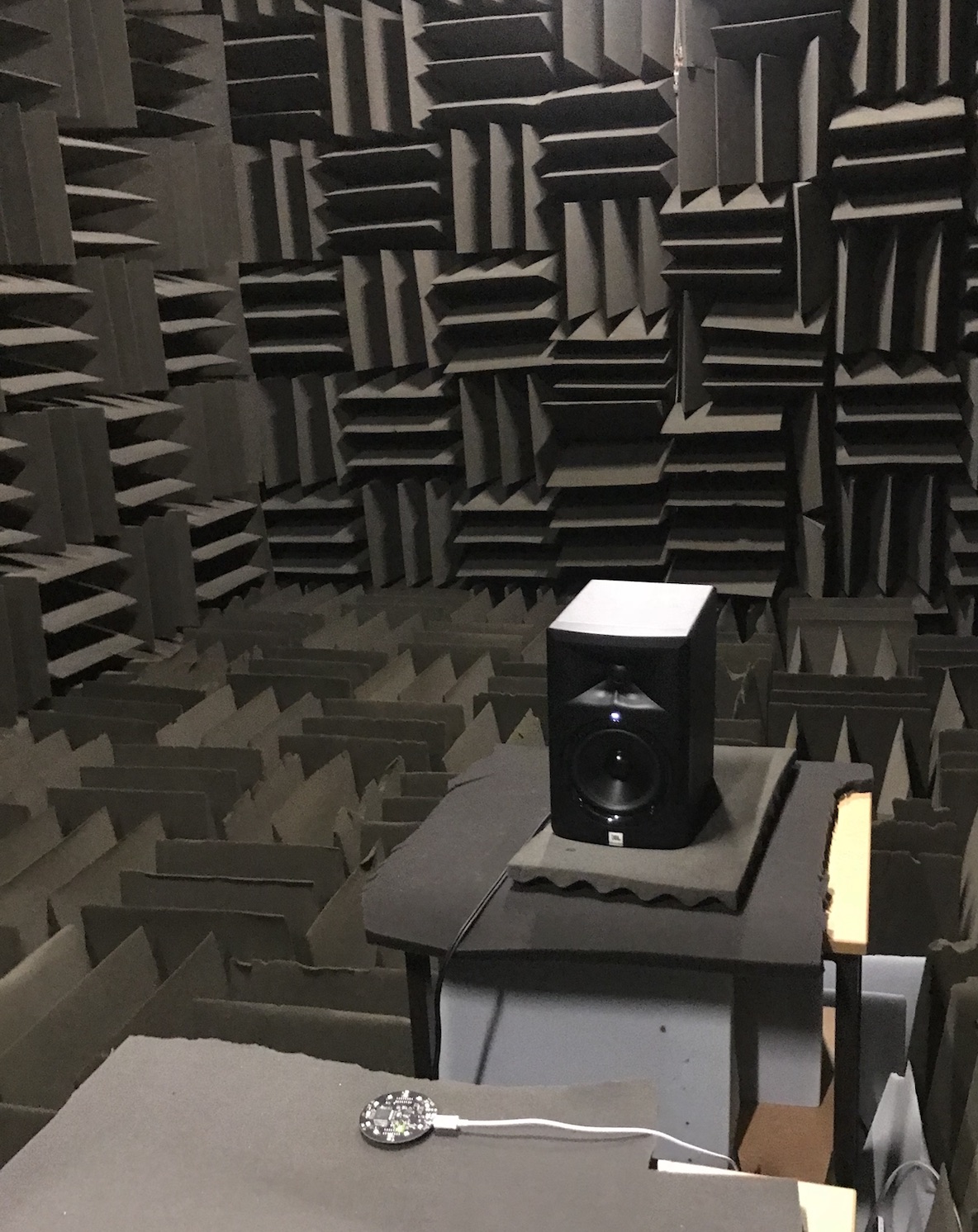}
\vspace{-0.4cm}
\caption{}
\end{subfigure}
\vspace{-0.4cm}
\caption{(a) Difference in mel-spectrograms of a speech segment as captured by three different microphones. The two rightmost figures show how data of Matrix Voice differs from ReSpeaker and PlugUSB. (b) Experiment setup in a non-reflective anechoic chamber to control for ambient acoustic variations.}
\label{diag.microphone_differences}
\vspace{-0.4cm}
\end{figure*}

Both the \emph{hardware} and \emph{software} components of this pipeline can introduce confounding artifacts in the audio signal. For example, different acoustic sensors are known to process audio signals differently, which introduces the first form of variability in the collected audio data. Das et al.\cite{das2014fingerprinting} showed that although these variabilities are subtle, they are computationally significant to even fingerprint smart devices based on their microphones. We argue that in the context of low-cost embedded devices, such hardware variabilities are likely to be even more prominent. 

On top of the hardware-related variations, the software processing pipelines of microphones can add further variabilities to the audio data. The DSP on embedded devices run a number of audio enhancement algorithms, whose parameters are fixed by each microphone manufacturer. For example, in the case of a microphone array, acoustic beamforming algorithms are employed to improve the SNR of the audio -- however, implementation differences in such algorithms across manufacturers can lead to variability in the output audio data.

{\parjump}

\noindent
\textbf{Embedded Microphone Variations.} 
We now provide empirical evidence of data variabilities in embedded microphones. We conduct an experiment in a non-reflective anechoic chamber (Figure~\ref{diag.microphone_differences}(b)) where the same audio signal is simultaneously recorded on three off-the-shelf embedded microphones, namely Matrix Voice, ReSpeaker, and a plug-and-play USB microphone (referred as PlugUSB). Matrix Voice and ReSpeaker are embedded microphone arrays designed for audio and speech applications, whereas PlugUSB is a single-channel microphone popular with devices such as Raspberry Pi. More details about these microphones are provided later in \S~\ref{subsec.method}.

In Figure~\ref{diag.microphone_differences} (a), we show mel-spectrograms of a 4-second speech segment as recorded by the three microphones simultaneously. We observe that the microphones exhibit differences in their frequency responses to the same speech input, which are also visualized in the two rightmost figures. For example, the data captured from Matrix Voice has low power in the high frequency ranges, which suggests that either this microphone does not capture high frequencies well (hardware effect) or they are being filtered out by the microphone DSP (software effect). Next, we discuss how these variabilities can impact the performance of audio classifiers. 

{\parjump}

\noindent
\textbf{Effect on audio classifiers.} 
From the perspective of audio models, the variability in hardware and software processing pipelines across microphones can be interpreted as a source of \emph{domain shift}. Domain shift refers to the phenomenon wherein the test data distribution differs from the training data distribution, leading to poor generalization performance of machine learning models. As such, if an audio model is trained on a certain type of microphone and tested on another type, the domain mismatch between the two microphones may cause degradation in model performance.

We now demonstrate how microphone variability can impact state-of-the-art speech models. We conduct an experiment in the context of an automatic speech recognition (ASR) task -- for this, we record the Librispeech-clean-test~\cite{panayotov2015librispeech} dataset on the three microphones presented earlier. Librispeech-clean-test is a benchmark dataset used to evaluate ASR model performance - it consists of 5.4 hours of U.S. English speech data from 40 speakers. The recordings are done simultaneously on all three microphones in a controlled environment with no background noise. We use Mozilla DeepSpeech2~\cite{mozilla} pre-trained ASR model as the target audio model upon which we evaluate the performance of the Librispeech datasets. This model has an error rate of 6.5\% on the original Librispeech-clean-test dataset. This model is trained on five different ASR datasets, none of which were collected from the embedded microphones on which we will evaluate the model. As such, this represents a scenario where there is a mismatch between the training and test microphones. 

\vspace{-0.2cm}
\begin{table}[htb]
\centering
\footnotesize
\begin{tabular}{|c|c|c|c|} \hline
\thead{Device} & Matrix Voice & ReSpeaker & PlugUSB \\ \hline
\thead{WER} & 41.13\% & 23.49\% & 26.10\% \\ \hline
\end{tabular}
\caption{Word error rate of pre-trained DeepSpeech2 on three embedded microphones.}
\vspace{-0.4cm}
\label{tbl:wer}
\end{table}
\vspace{-0.2cm}

Table~\ref{tbl:wer} shows the word error rates (WERs) of DeepSpeech2 on each microphone-specific test set. We observe that: a) although the DeepSpeech2 model has an advertised WER of 6.5\% on the original Librispeech-clean dataset, when we deploy it on different embedded microphones, the WER increases above 20\%. b) more importantly, we observe that the WER varies significantly across different microphones (ranging from 23.49\% to 41.13\%). These two initial findings highlight that the mismatch between training and test microphones has a severe impact on audio models. Particularly in the context of embedded devices with high likelihood of microphone variability, it is important to tackle this challenge to build robust and usable audio systems.

{\parjump}

\noindent
\textbf{Existing Solutions.} 
A simple solution to this problem could be to train an audio model on data collected from a large set of microphones such that the model learns features which are invariant to microphone variations. While possible, we argue that it is an expensive and non-scalable solution because it involves collecting real-world training data on multiple microphones. More critically, new wearable and embedded devices are being released at a rapid pace, often with integrated custom microphones. Therefore, collecting large-scale audio datasets from such diverse and newly emerging devices is not feasible. 

Recently, multichannel audio modeling techniques~\cite{sainath2017multichannel, li2017acoustic} have been proposed with the goal of improving ASR accuracy on microphone arrays. Operating on the raw waveforms from the microphone hardware, these works aim to jointly optimize the parameters of the signal processing algorithms (e.g., source localization, beamforming) with those of the audio model~\cite{seltzer2004likelihood}. While these approaches can minimize the signal variability caused by different DSPs, they are not ideal for embedded devices for two reasons: a) models based on raw audios are computationally expensive~\cite{sainath2017multichannel} for resource-constrained devices; b) they make an assumption that the microphone hardware and microphone array geometry remains identical between the training and test runs, which may not hold for the diverse market of embedded devices. 

Finally, domain generalization techniques have been proposed to learn noise-invariant feature representations for speech models~\cite{2018arXiv180706610L, serdyuk2016invariant}, however these techniques rely on paired data (i.e., pairs of noisy and clean speech). While constructing large number of paired noisy samples from clean speech is a trivial task, in the case of microphone variability, it is very expensive to get paired samples from different microphones. Similarly, \cite{ipsn18_mathur} presented a data augmentation solution for improving the robustness of audio models on smartphones and smartwatches, however their solution also assumes the availability of aligned samples from multiple microphones, which as we argued is not scalable in the real-world. Ideally a solution which can be designed just using unpaired data from different microphones would be considered more practical.

{\secspace}
\section{Mic2Mic Overview} 
\label{sec:overview}
{\secspace}
In this section, we provide an overview of our solution to address microphone variability. Our primary goal is to develop a \emph{practical} solution which can generalize to multiple audio tasks and which does not put additional burden on the end-users to provide labeled training data from new microphones. 

\subsection{Design Considerations}
\label{design}
The following four key considerations have shaped the design of our solution, \system{}.

\squishlist{

\item{\textbf{Decouple microphone variability problem from the audio task.} Domain shift caused by the underlying microphone variability is independent of the task-specific audio model (e.g., ASR, keyword detection). As such, it is important for generalizability that \system{} is independent of the downstream audio task and hence can be applied to diverse tasks. Moreover, many commercial audio models are proprietary. As such, while developing and deploying \system{}, we should not assume that parameters of the task-specific model are known.}

\item{\textbf{Avoid re-training of the task-specific audio model.} In continuation of the previous design consideration, \system{} should adapt to unseen microphones without requiring any retraining of the task-specific audio model.}

\item{\textbf{Minimize burden on the end-users: } When \system{} is to be deployed on an unseen microphone (or device) in the real-world, it should not burden the user with providing carefully calibrated and labeled data. Ideally, the solution should be developed using unconstrained and unlabeled audio data.}

\item{\textbf{Plug-and-play system component:}  \system{} should provide an abstraction of a plug-and-play system component that system developers can simply import in their inference pipeline, independent of the audio task.}
}

\squishend{}

\begin{figure}[t]
\centering
\includegraphics[width=0.6\linewidth]{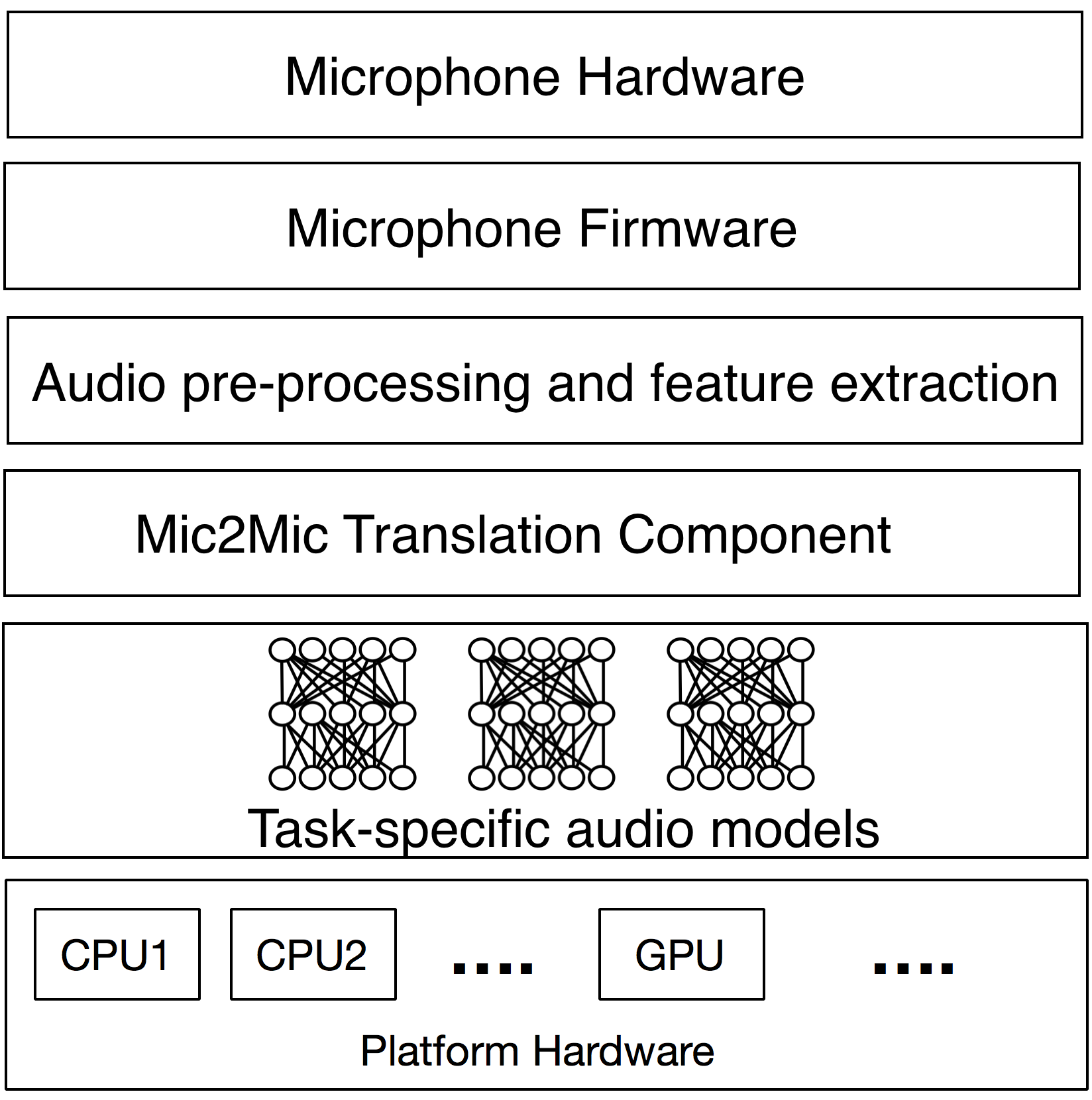}
\vspace{-0.3cm}%
\caption{System Overview of Mic2Mic}
\vspace{-0.6cm}
\label{diag:arch}
\end{figure}

\subsection{Architecture and Dataflow}
Before discussing how Mic2Mic is trained, we briefly describe how it will be incorporated in an embedded system to solve for microphone variabilities. Figure ~\ref{diag:arch} depicts the overall architecture of our system, and its main components are as follows:

\squishlist{
\item{\textbf{Microphone hardware and firmware.} The microphone hardware captures the analog audio signal and digitizes it using an ADC to output PCM values. Optionally, in the case where an embedded device consists of multiple microphones (e.g., a microphone array), the output of each individual microphone is combined using software processing techniques such as delay-and-sum beamforming. Both the hardware and software components of a microphone pipeline can introduce variabilities in the audio data, thereby leading to the \emph{domain shift} problem.} 

\item{\textbf{Preprocessing and feature extraction.} Digitized audio data exposed by the microphone's software framework is first preprocessed (e.g., audio segmentation, volume normalization), and optionally light features are extracted from the audios. Our current implementation uses log-spectrogram features, however other kinds of features can be easily incoporated into the framework.} 

\item{\textbf{Mic2Mic Translation Component.} Mic2Mic is a machine-learned system component to minimize the domain shift caused by microphone variability. Developers import Mic2Mic in their inference pipeline, pass the log-spectrogram features as input to it and specify the target microphone. Mic2Mic then performs a real-time \emph{translation} of the audio data from current microphone (or test microphone) to the target microphone (or training microphone). In other words, Mic2Mic pushes the test data towards the training data distribution.}

\item{\textbf{Audio Model.} Finally, the translated audio data is passed to the task-specific audio model (e.g., ASR or keyword detection) for computing inferences. As the translated data is closer to the training data distribution than the original test data, we expect the inference accuracy to increase.}
}

\squishend{}

{\secspace}
\section{Formulating A Generative Modeling Solution} 
\label{sec:approach}
{\secspace}
In this section we describe how Mic2Mic is trained using unlabeled, unpaired data from multiple microphones. We first provide a primer on generative adversarial networks (GANs), then discuss how the problem of microphone variability can be formulated as a GAN translation problem.

\subsection{Primer on generative adversarial networks}
\label{sec:gan_intro}
Deep learning has enabled many new applications of discriminative modeling, that is learning to predict a label $y$ for an input $x$.
On the other hand, generative modeling means learning to draw samples from a distribution $p(x)$, or in the presence of labels, from $p(x, y)$.
For example, in a keyword detection model, the task is to predict the probability of the presence of a keyword in an audio segment, and it is achieved in a \emph{discriminative} fashion by choosing a keyword class with the highest output probability for the given audio segment. On the other hand, a generative modeling task here could be to \emph{generate} an audio segment containing a given keyword class.

Recent work using generative adversarial networks (GANs) is focused on extending deep learning techniques to work in the generative context \cite{goodfellow2014generative}.
The difficulty in conventional generative modeling techniques is the lack of an obvious evaluation metric, which can measure the `goodness' of the generated data. 
GANs solve this problem using two neural networks $G$ and $D$, the generator and the discriminator, respectively. The generator $G$ takes a noise vector $z$ as input and generates a data sample by evaluating $G(z)$. Besides noise, other information can be fed into the generator, in which case $G$ is called a conditional generator. The discriminator $D$ on the other hand is trained to distinguish between the real samples from $p(x)$ and the generated samples from G. Effectively, the discriminator provides feedback about the `goodness' of the generated sample to G, which uses this feedback to generate even better data samples and fool the discriminator. In this way, the two neural networks $G$ and $D$ play a competitive game and in the process, both become better at their respective tasks: the generator ends up generating high-quality data that resembles the distribution $p(x)$, and discriminator becomes good at distinguishing data drawn from $p(x)$ vs. other data distributions.

{\parjump}
\noindent
\textbf{GANs for data translation}. GANs have been recently used for the task of data translation, particularly with images. Assume we want to learn a mapping or translation between two image domains, namely colored and black-and-white (B/W). The GAN takes as input a \emph{paired} set of images $(a, b)$ from the two domains where $a$ is a colored image and $b$ is the corresponding B/W image. The colored image $a$ is fed to the generator, and the output $G(a)$ is compared against the paired B/W image $b$ by the discriminator. D provides feedback to G about the `goodness' of the generated B/W samples, and G uses this information to learn an even better mapping between colored and B/W image domains. Recent works\cite{isola2017image} have shown remarkable results in \emph{paired} image data translation using conditional GANs.

\subsection{Microphone Variability as a Translation Problem}
\label{sec:motivation}
We propose to formulate the problem of microphone variability as a data translation problem, i.e., given an audio from a microphone (e.g., a test microphone), can we translate it to a different microphone's (e.g., a training microphone) domain? If a translation function can indeed be learned between training and test microphones, it can subsequently be used to reduce the domain shift caused by microphone variability.

A major challenge however is that generating large-scale paired and aligned audio datasets from multiple microphones is not trivial. This would require asking end-users to provide pre-specified speech inputs from their microphones (e.g., repeating 'Hey Siri' 100 times) and moreover these inputs would need to be carefully time-aligned to create a paired dataset of multiple microphone audios upon which a conditional GAN can be trained. As mentioned in \S~\ref{design}, a key design consideration for Mic2Mic is to minimize the burden on end-users, as such the above approach is not ideal. 

Therefore, we seek a solution which can learn the mapping between two microphone domains using just \emph{unpaired} and \emph{unlabeled} data. In this work, we adopt principles of the CycleGAN posed in \cite{zhu2017unpaired} to learn an audio translation model using unpaired data. 

{\parjump}
\noindent
\textbf{Mic-to-Mic translations using CycleGAN}. Assume we have a set of unpaired speech samples from two microphones $A$ and $B$ and we wish to learn a mapping $G_{A\rightarrow B}$. As discussed in \S\ref{sec:gan_intro}, a GAN can be trained such that it outputs a distribution $G_{A\rightarrow B}(A)$ which is indistinguishable from the data distribution of B. However, it does not guarantee that each individual speech sample $a_{i}$ from the source microphone $A$ is mapped to its corresponding output $b_{i}$ in the target microphone $B$ -- in other words, it is not guaranteed that the speech content of the input sample will be preserved by the mapping $G_{A\rightarrow B}(a)$. In order to solve this, CycleGAN imposes a cycle-consistency structure on the mappings, in that if an input sample $a$ is translated from $A\rightarrow B$ and then back from $B\rightarrow A$, we should arrive at the same input sample. As such, CycleGAN proposes to learn two bijective mapping functions (or generators) $G_{A\rightarrow B}$ and $G_{B\rightarrow A}$ such that both are inverse of each other. By imposing the cycle consistency structure, CycleGAN is able to learn one-to-one mappings between the source and target domains. In Section~\ref{sec:cycle}, we provide details on the architecture of our CycleGAN model, and how we train and deploy it on an embedded device.

{\secspace}
\section{Mic2Mic Training and Implementation}
\label{sec:opt}
{\secspace}
In this section, we explain how the Mic2Mic translation model is trained and then implemented on an embedded device as a system component. 

\begin{figure}[htb]
\centering
	\includegraphics[width=\linewidth]{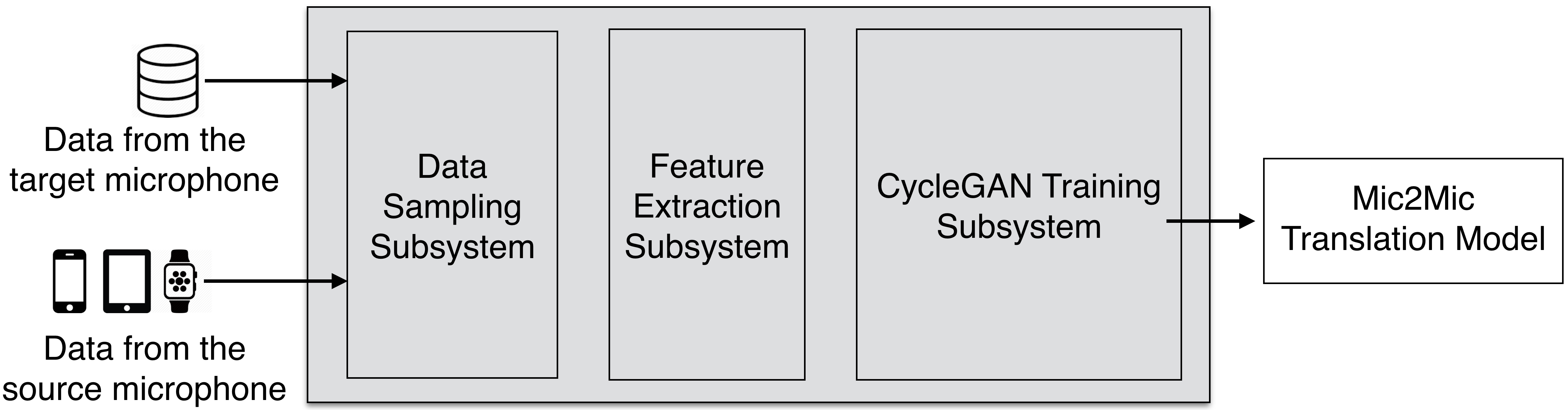}
	\vspace{-0.2cm}
	\caption{Mic2Mic training system}
	\label{diag:training}
	\vspace{-0.2cm}
\end{figure}

\subsection{Training of Mic2Mic}
\label{sec:cycle}
In Figure ~\ref{diag:training}, we show the architecture of Mic2Mic's training system. In our current implementation, the training of the translation model is done centrally on the cloud. Below are the main components of the training system. 

\parjump{}
\noindent{}
\textbf{Data Collection Subsystem.} To train a translation model, we require data from source and target microphones. The source microphone here refers to the deployment microphone on the embedded device, whereas the target microphone is a representative microphone that was used for collecting the training data for the task-specific audio model. We assume that the target microphone is known and unlabeled data from it is available (e.g., it could be provided by the model developer or available in a public dataset).

For the source microphone, the Data Collection subsystem requests speech data from the embedded device. We envision a system component called Training Manager (described in \S~\ref{sec:inference}) on the user's device which -- on receiving a data request from the Data Collection subsystem -- is able to collect speech data from the embedded (source) microphone with user's permission. As Mic2Mic only requires unpaired and unlabeled data, it is very cheap to collect without requiring extensive effort from end-users.

\parjump{}
\noindent{}
\textbf{Feature Extraction Subsystem.} Once the source and target microphone datasets are collected, the raw audios are converted to log-spectrograms. To compute log-spectrograms, we use a sliding-window approach with a Hamming window size of 32 ms and a hop size of 16 ms, with FFT bin width of 256. The log-spectrograms are normalized to be in the range $[-1, 1]$. Although our current implementation operates over log spectrograms, in principle it can be extended to other audio features. We denote the unaligned empirical distributions of log-spectrograms from the source and target microphones as $p(x_A)$ and $p(x_B)$, respectively.

\begin{figure}[htb]
\begin{subfigure}[b]{0.49\linewidth}
\includegraphics[width=0.7\linewidth]{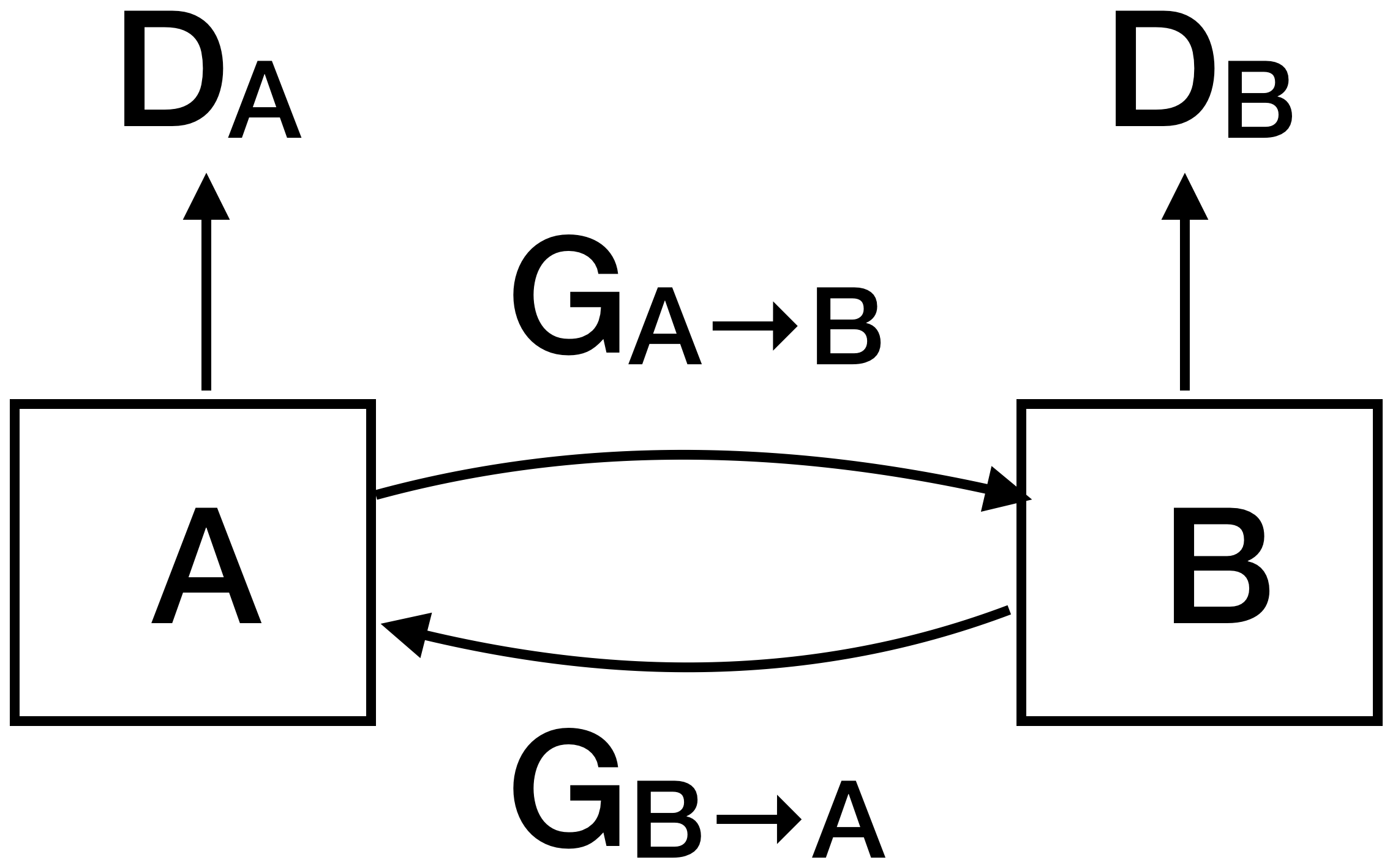}
\label{cyclegan1}
\vspace{-0.4cm}
\caption{ }
\end{subfigure}
\begin{subfigure}[b]{0.49\linewidth}
\includegraphics[width=0.8\linewidth]{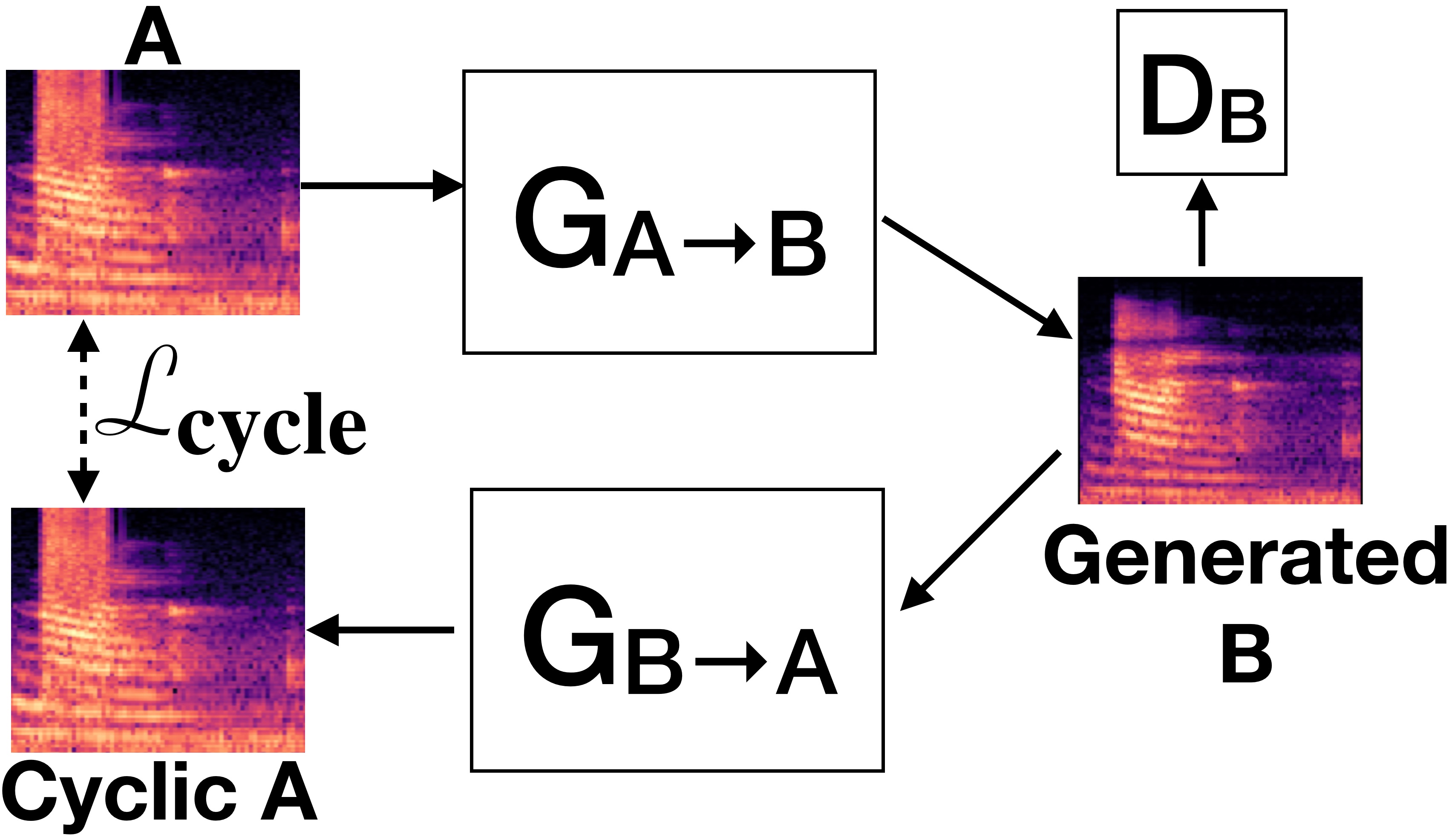}
\label{cyclegan2}
\vspace{-0.4cm}
\caption{ }
\end{subfigure}
\vspace{-0.4cm}
\caption{(a) Architecture of a cycleGAN consisting of two bijective generators and two discriminators. (b) Illustration of the Cycle Consistency property --  minimizing $\mathcal{L}_{cycle}$ helps in preserving the content of the translated audio.}
\label{cyclegan}
\vspace{-0.4cm}
\end{figure}

\parjump{}
\noindent{}
\textbf{CycleGAN Training Subsystem.} We now explain how we learn the mapping between $p(x_A)$ and $p(x_B)$ using a CycleGAN. As shown in Figure~\ref{cyclegan}, a CycleGAN architecture consists of two generators $G_{A\rightarrow B}$ and $G_{B\rightarrow A}$ and two corresponding discriminators $D_{B}$ and $D_{A}$.  Our two \emph{generator} models follow the U-Net architecture\cite{ronneberger2015u} and consist of 3 convolutional layers for extracting higher level features from the spectrograms, 3 ResNet blocks for transforming the features from source domain to target domain, followed by 3 transpose convolutional layers for converting the transformed features into output spectrograms. We add skip connections\cite{ronneberger2015u} between the convolutional and corresponding transpose convolutional layers and use batch normalization layers between the ResNet blocks for faster convergence. For the \emph{discriminator} models, we use a 4-layer deep fully-convolutional network with a batch normalization layer between two consecutive layers. 

The training pipeline for the CycleGAN works as follows. Samples from the distribution $p(x_A)$ are fed to a generator $G_{A\rightarrow B}$, and the outputs $G_{A\rightarrow B}(x_A)$ are evaluated using a discriminator $D_B$, which compares them to actual samples from $p(x_B)$. The loss function used for training the CycleGAN is composed of three terms. For the generator $G_{A\rightarrow B}$, they are as follows: 

{\parjump}
\textbf{Least square generator loss.} It measures how distinguishable the generated data $G_{A\rightarrow B}(x_A)$ are from the target data distribution $p(x_B)$. As the distribution of the generated data becomes closer to $p(x_B)$, this loss becomes smaller. Mathematically, it is expressed as:

   \begin{equation}
     \mathcal{L}_{adv}(G_{A\rightarrow B}) = \mathbb{E}_{x\sim p(x_A)}\left[(D_B(G_{A\rightarrow B}(x)) - 1)^2\right]
   \end{equation} 

\textbf{Cycle Consistency Loss.} This term enforces the cycle consistency property and drives the two generators to be inverses of one another. 

    \begin{equation}
      \mathcal{L}_{cycle}(G_{A\rightarrow B})= \mathbb{E}_{x\sim p(x_A)}\left[\|G_{B\rightarrow A}(G_{A\rightarrow B}(x)) - x\|_{L_1}\right]
    \end{equation}

\textbf{Identity Loss.} This loss term drives the generators to be close to identity on samples drawn from the target distribution.

    \begin{equation}
      \mathcal{L}_{id}(G_{A\rightarrow B}) = \mathbb{E}_{x\sim p(x_B)}\left[\|G_{A\rightarrow B}(x) - x\|_{L_1}\right]
    \end{equation}

The total loss which is minimized for training the generator $G_{A\rightarrow B}$ takes the form: 
\begin{equation}
  \mathcal{L}_{G_{A\rightarrow B}} = \alpha\mathcal{L}_{adv} + \beta\mathcal{L}_{cycle} + \gamma\mathcal{L}_{adv},
\end{equation}
where the coefficients $\alpha, \beta, \gamma>0$ can be adjusted as hyper-parameters.

The discriminator $D_{B}$ is trained by minimizing the least squares discriminator loss:

\begin{equation}
  \mathcal{L}_{D_B} = \mathbb{E}_{x\sim p(x_A)}\left[(D_B(G_{A\rightarrow B}(x)))^2\right] + \mathbb{E}_{x\sim p(x_B)}\left[(D_B(x) - 1)^2\right].
\end{equation}

The pair $(D_A, G_{B\rightarrow A})$ is trained similarly, with subscripts $A$ and $B$ interchanged. After the cycleGAN is trained, we are primarily interested in the (source$\rightarrow$ target) generator $G_{A\rightarrow B}$, which can take audio features from the source microphone and generate equivalent audio features from the target microphone. As such, we discard the (target$\rightarrow$source) generator $G_{B\rightarrow A}$ and the two discriminators. 

\begin{figure}[htb]
\centering
	\includegraphics[width=\linewidth]{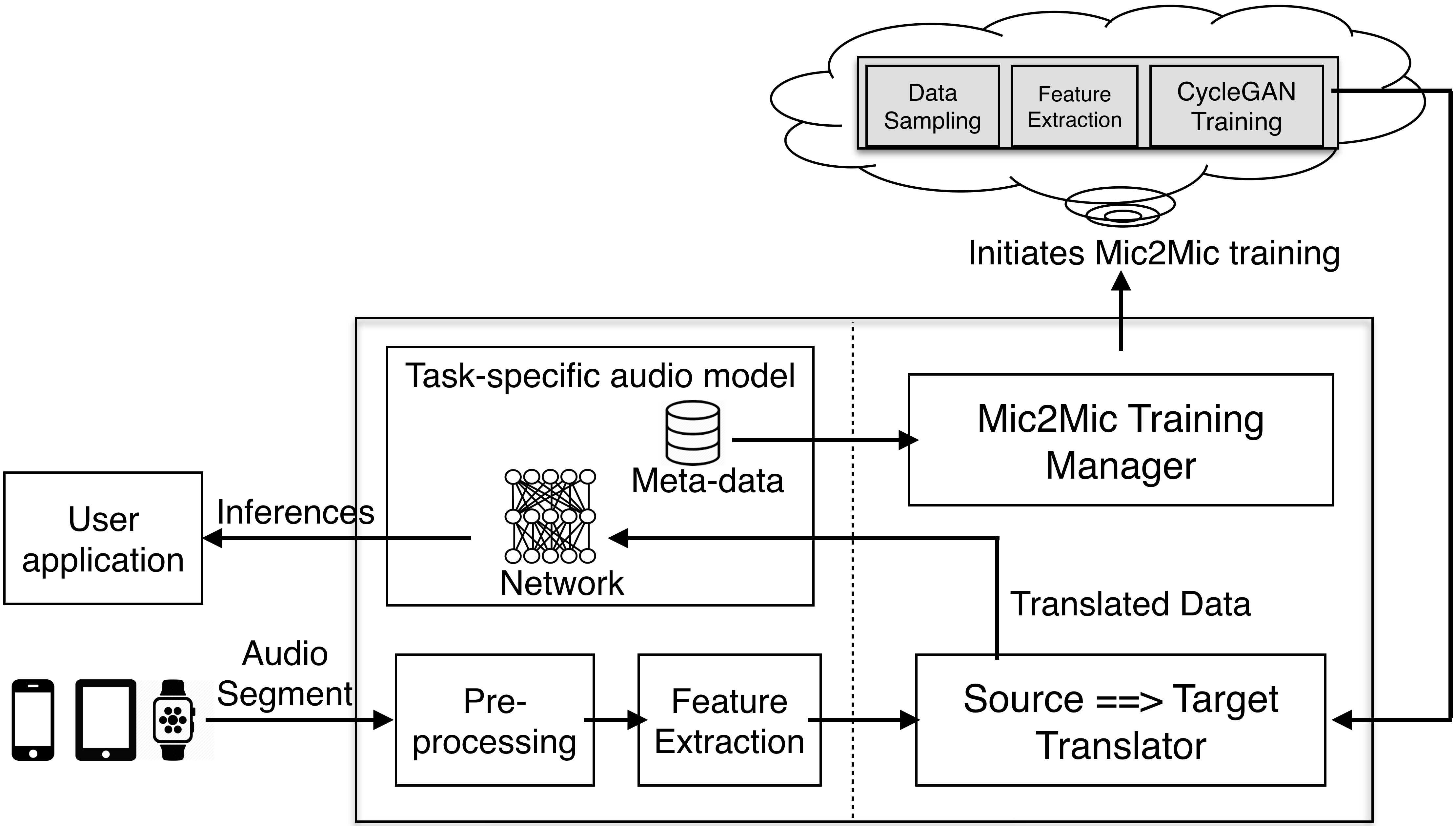}
	\vspace{-0.4cm}
	\caption{System diagram of Mic2Mic deployment}
	\label{diag:inference}
	\vspace{-0.4cm}
\end{figure}

\subsection{Mic2Mic Deployment}
\label{sec:inference}

In this section, we discuss the implementation of Mic2Mic as a system component on embedded devices. Figure~\ref{diag:inference} shows the system architecture and its main components are as follows:

\parjump{}
\noindent{}
\textbf{Training Manager.} This sub-component is responsible for initiating the training of Mic2Mic's translation model. We assume that task-specific audio model (e.g., ASR) provides information about its training microphone(s) in the form of meta-data. The training manager reads this metadata and decides if a translation model needs to be learned. For instance, if the deployment microphone (i.e., the microphone on the embedded device) is the same as the training microphone, the problem of domain shift due to microphone variability does not arise, as such a translation model is not needed. Alternatively, if the deployment  microphone differs from the training microphone, the \emph{training manager} initiates training of the translation model.

As described in \S~\ref{sec:cycle} and Figure~\ref{diag:training}, the training process currently runs on the cloud and is dependent on receiving unlabeled speech data from the deployment microphone. The training manager initiates the collection of this data by prompting the user to provide speech inputs on the deployment microphone. Upon receiving the user input, we first filter it through a voice activity detector (VAD) module to separate speech content from other audio data. The segmented speech content is then uploaded to the cloud to start Mic2Mic's training process. 

Note that in this paper, we do not conduct a real-world user study wherein the speech data is collected from the users directly. Instead, as we will describe in \S~\ref{subsec.method}, we use a pre-recorded speech dataset as a proxy for a user, and the training manager samples speech from this dataset. Future work can extend our work by conducting a user study in which speech data is incrementally collected from the users. Moreover, other ways of collecting speech data such as piggybacking on user's speech inputs to various applications (with user's permission) could be explored.

\parjump{}
\noindent{}
\textbf{Mic2Mic Translator.} The output of the training process is a (source$\rightarrow$target) translation model which is downloaded on the embedded device. Thereafter, upon receiving audio data from the device microphone, the system pre-processes the data and computes the log-spectrogram features required by the translation model. The computed features from the source domain (i.e., device microphone) are then translated to the target domain (i.e., the training microphone) in order to reduce the domain shift. Finally, the translated features are fed to the task-specific audio model for computing inferences. As many audio models (e.g., keyword detection~\cite{zhang2017hello}, DeepSpeech2~\cite{mozilla}) use spectrograms or MFCC coefficients as input features, they are compatible with our current implementation.

{\secspace}
\section{Evaluation}
\label{sec:eval}
{\secspace}
\label{sec:eval}
\noindent
In this section, we present a systematic evaluation to highlight the accuracy gains achieved by \system{}. The key research questions that drive our experiment are:

 \squishlist {
 \item{Does incorporating \system{} in the inference pipeline improve an audio model's accuracy on unseen microphones (`unseen' refer to those microphones which the model did not `see' during the training process)?} 
 \item{How does the performance of \system{} compare against audio preprocessing and microphone calibration techniques?}  
\item{How much unpaired audio data is needed to train the \system{} translation model? }
\item{What is the system overhead of running the translation model on embedded devices?}
 }
 \squishend
 {\parjump}

\noindent
The key highlights from our experimental results include:
 \squishlist {
 \item{\system{} is able to effectively learn a microphone translation function using less than 20 minutes of unlabeled and unpaired data.} 
 \item{Mic2Mic can recover between 67\% to 89\% of the accuracy lost due to microphone variability for two common audio tasks.}
 \item{It is feasible to run Mic2Mic on mobile and embedded devices within reasonable resource contraints.}
 }
 \squishend
 {\parjump}

{\subsecspace}
\subsection{Data Collection}
{\subsecspace}
\label{subsec.method}
\noindent
Our goal is to evaluate the performance of \system{} in improving the robustness of audio models when they are deployed on microphones different from those used while training. For this, we need two types of datasets: a) dataset for training and testing task-specific audio models, and b) dataset for training and testing Mic2Mic's translation model. More importantly, these datasets should be collected from different types of embedded microphones, while at the same time controlling for confounding factors such as ambient noise and speech content. To the best of our knowledge, no such datasets exist in the public domain and therefore, we first present our methodology to create these datasets.  

\begin{figure}
  \begin{minipage}[b]{0.49\linewidth}
    \centering
\footnotesize
\begin{tabular}[b]{|c|c|c|c|} \hline
\thead{Device} & \thead{Count} & \thead{Cost} \\ \hline
Matrix Voice & 2 & \$55 \\\hline
ReSpeaker  & 2 & \$80  \\\hline
PlugUSB  & 2 & \$5 \\\hline
\end{tabular}

  \end{minipage}
  \hfill
  \begin{minipage}[b]{0.15\linewidth}
    \centering
   \includegraphics[width=\linewidth]{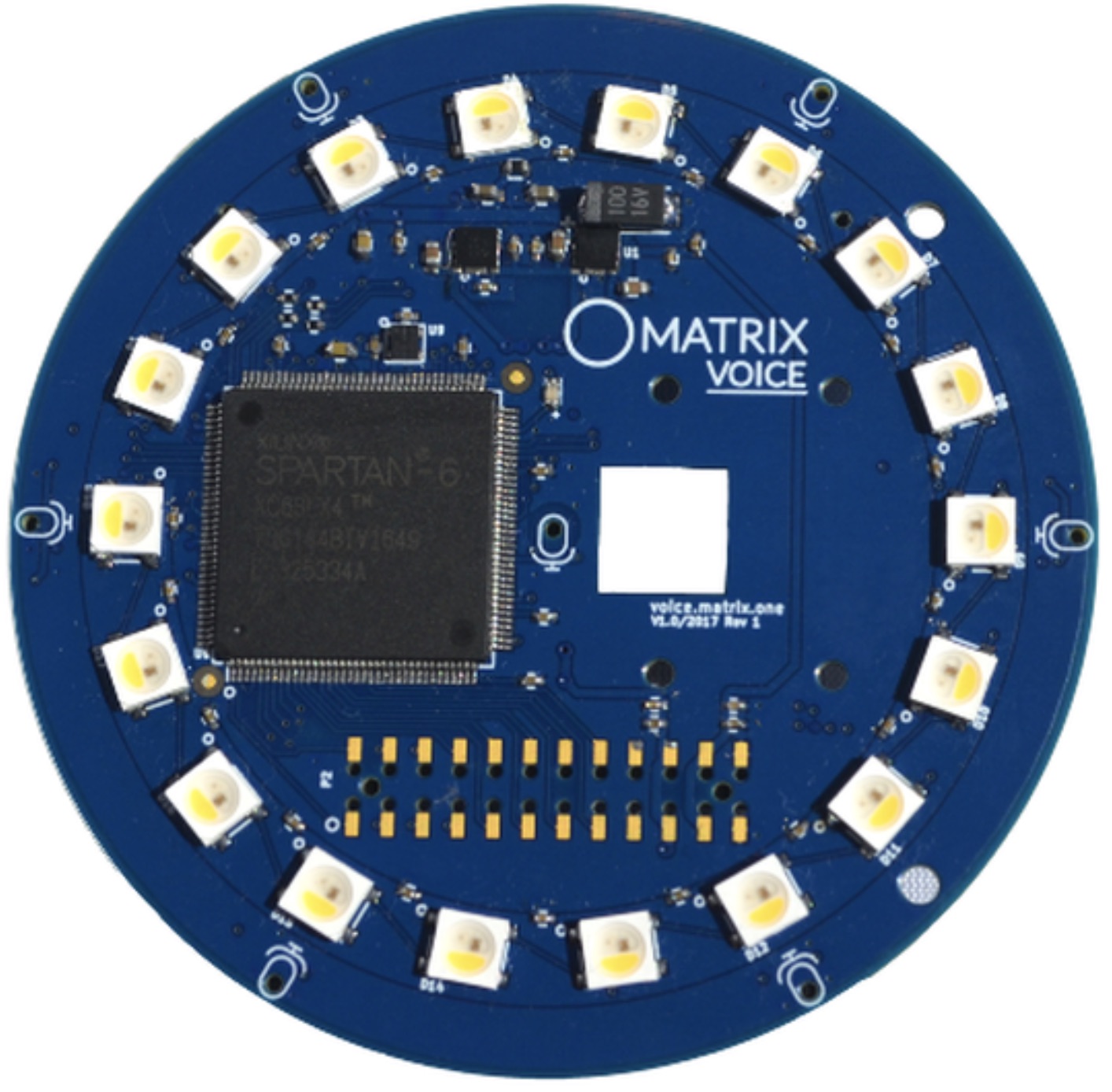}
    \end{minipage}
  \begin{minipage}[b]{0.15\linewidth}
    \centering
   \includegraphics[width=\linewidth]{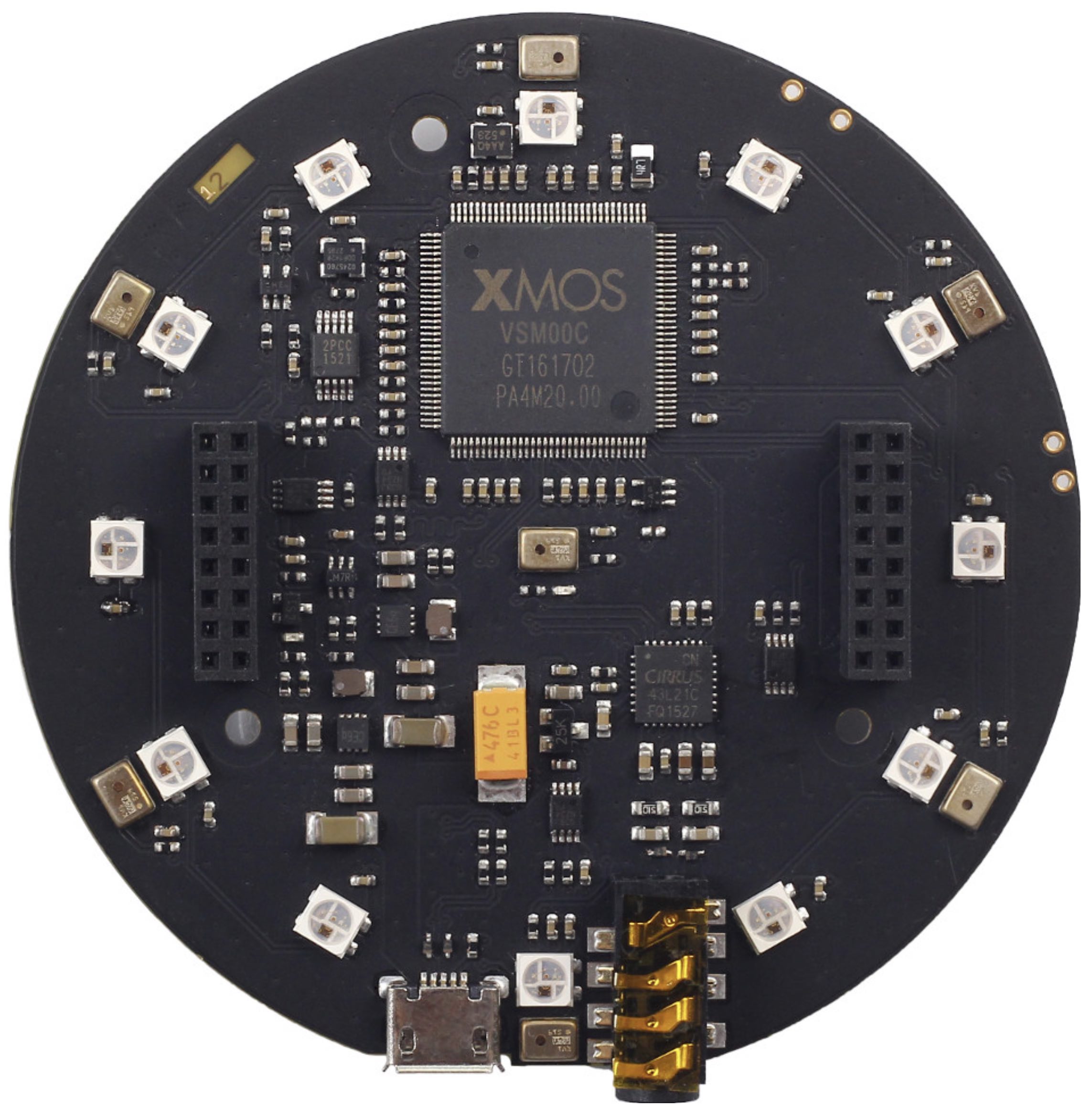}

    \end{minipage}
  \begin{minipage}[b]{0.15\linewidth}
    \centering
   \includegraphics[width=0.7\linewidth]{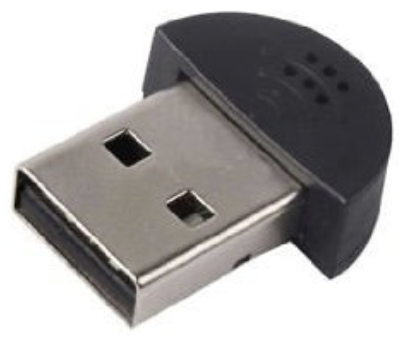}
    \end{minipage}
\caption{Microphones used in our experiments. (Left to right) Matrix, ReSpeaker, PlugUSB}
\vspace{-0.5cm}
\label{fig.devices_audio}
\end{figure}

{\parjump}
\noindent
\textbf{Microphone Hardware.} %
We collect audio data from six different microphones representing three class of devices. As shown in  Figure~\ref{fig.devices_audio}, we use two circular microphone arrays, namely Matrix Voice (costing \$55) and ReSpeaker (costing \$80). Both are programmable microphone arrays consisting of seven MEMS microphones located on the periphery of the device. They record audios from all 7 microphones and perform on-device signal processing such as delay-and-sum beamforming to enhance the audio signal. Eventually, they generate a single processed audio which represents the best quality audio recording from these devices. The third class of microphones used in our experiments are low-cost single-channel USB microphones (costing \$5). In effect, these devices represent a range of microphones that are commercially available for developing speech applications on embedded platforms. We use a Raspberry Pi 3 Model B as our target embedded platform because all three classes of microphones are compatible with it.

{\parjump}

\noindent
\textbf{Audio Tasks and Datasets.} %
Two representative audio sensing tasks are used in our experiments, namely Keyword Spotting and Emotion Detection.  

{\parjump}
\noindent
\textit{Keyword Spotting.} %
In this task, the goal is to identify the presence of a certain keyword class (e.g., Hey Alexa) in a given speech segment. To train a model for this task, we use the \emph{Speech Commands} dataset containing 65,000 one-second long utterances of 30 short keywords~\cite{speechcommands}. Instead of using all 30 classes, we used a subset of 12 classes for our experiments (yes, no, up, down, left, right, on, off, stop, go, zero, one). This 12-class dataset (referred to as \emph{SC-12}) was then randomly split into training (75\%) and test (25\%) class-balanced subsets. We use a small-footprint keyword detection architecture proposed in ~\cite{zhang2017hello} to train the model. The input to this model is a two-dimensional tensor extracted from the one-second long keyword recording, consisting of time frames on one axis and 24 MFCC features on the other axis. The model outputs a probability of a given audio recording belonging to a certain keyword class (e.g., Yes, No) or to an Unknown class.

{\parjump}
\noindent
\textit{Emotion Detection.} In this task, the goal is to identify the emotion of the speaker in a given speech segment. To train a model for this task, we use the \emph{RAVDESS} dataset which is a collection of 1440 speech files recorded by 24 actors where they expressed a range of emotions such as calm, happy, sad, angry, fearful, surprise, and disgust. The dataset was randomly split into training (75\%) and test (25\%) class-balanced subsets. We use a CNN-based speech emotion detection architecture proposed by \cite{badshah2017speech} for training our models.

{\parjump}

\noindent
\textbf{\system{}  training data.} %
In addition to the task-specific models, we also need data to train Mic2Mic translation models. As discussed, from the 30-class Speech Commands dataset, we only used data from 12 keyword classes (\emph{SC-12}) to train the keyword detection model. Rest of the data (\emph{SC-rest}) was used to train the Mic2Mic translation model. Note that there is no data overlap between \emph{SC-12} and \emph{SC-rest}, as such the translation model is trained completely independent of the task-specific model. 

\begin{figure*}[t]
\captionsetup[subfigure]{labelformat=empty}
\centering
\begin{subfigure}[b]{0.49\linewidth}
\includegraphics[width=\linewidth]{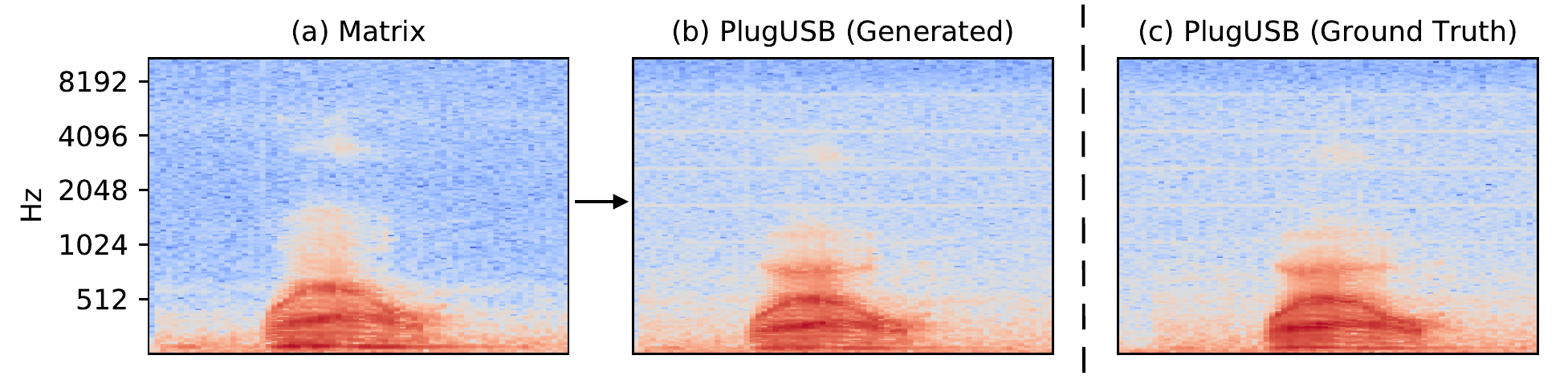}
\label{matrix-translation}
\vspace{-0.5cm}
\caption{\scriptsize Matrix$\rightarrow$PlugUSB Translation}
\end{subfigure}
\begin{subfigure}[b]{0.49\linewidth}
\includegraphics[width=\linewidth]{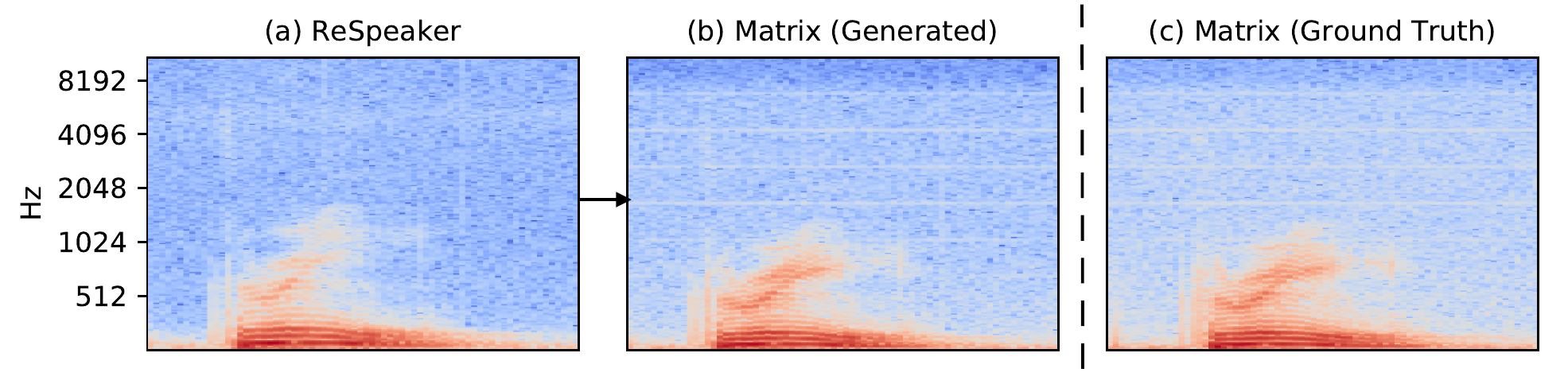}
\label{rs-translation}
\vspace{-0.5cm}
\caption{\scriptsize ReSpeaker$\rightarrow$Matrix Translation}
\end{subfigure}
\vspace{-0.4cm}
\caption{Translation Performance of \system{} for two different microphone pairs. (a) shows an example spectrogram from Microphone A and (b) shows the spectrogram generated by applying  $G_{A\rightarrow B}$ translation on (a). (c) shows the corresponding ground truth spectrogram for microphone B. We observe that the generated spectrograms (b) visually resemble their corresponding ground truth (c).}
\label{translations}
\vspace{-0.4cm}
\end{figure*}

{\parjump}

\noindent
\textbf{Setup.} The three datasets described above have a combined duration of around 20 hours and need to be recorded on all six microphones in our experiment. At the same time, we need to control for factors such as variability in the acoustic environments and speakers so that the only difference between the datasets are due to the microphone variability. Given these requirements and the large amount of data to be collected, we decided not to use human subjects for recordings. Instead, we use a JBL LSR305 reference monitor speaker to replay the audio datasets in a quiet room. This speaker has a relatively flat frequency response in the human speech range which allows for a faithful replay of the datasets. The replayed audios are recorded simultaneously on 6 Raspberry Pi devices, each connected with a microphone and placed equidistant (12cm) from the audio source. Effectively, we created six versions of the audio datasets, one for each microphone used in our experiment. 

Note that the data samples are collected in a `paired' manner only to facilitate the evaluation of our method -- the training of Mic2Mic does not rely on paired or labeled data and as we will discuss later, we enforce that no paired samples from different microphones are available for training Mic2Mic.

{\subsecspace}
\subsection{Evaluation of the translation model}
{\subsecspace}
\label{subsec.cyclegan}
\noindent
For evaluating the translation model, we use an `aligned' dataset of 1000 speech segments from two microphones. That is, we record the same speech segment with two different microphones and time-align them. This way we have a ground truth to compare the quality of our translations. As our translation model operates on 2D spectrograms, we use the image similarity metric of Peak Signal-to-noise ratio (PSNR) to evaluate the performance of the translation model. PSNR is closely related to the mean square error and is suggestive of the distance between two images. The higher the PSNR, the closer the images are to each other. 

\noindent
\textbf{Results.} %
We first present qualitative results to demonstrate the performance of \system{}. Figure~\ref{translations} shows the performance of two translation models: Matrix$\rightarrow$ PlugUSB and ReSpeaker $\rightarrow$ Matrix. Spectrograms $a$ and $c$ correspond to speech segments collected from source and target microphones, whereas the spectrograms $b$ are generated by applying $G_{A\rightarrow B}$ translation model on $a$. We observe that the generated spectrograms (b) visually resemble their corresponding ground truth (c), thereby suggesting that \system{} was able to learn a translation function between the source and target microphones. 

In Table~\ref{eg-table}, we show the effect of \system{} on the PSNR between training and test spectrograms. We observe that there is a significant increase in the PSNR pre- and post- translation, which suggests that Mic2Mic is able to reduce the domain shift between the microphones. 

\begin{table}[htb]
\captionsetup{font=small}
\centering
\begin{small}
\begin{tabular}{lccc}
\toprule
{\parbox{1.5cm}{\centering Training\\Microphones}} & {\parbox{1.5cm}{\centering Test\\Microphones}} & {\parbox{1.5cm}{\centering Untranslated\\PSNR}} &  {\parbox{1cm}{\centering Translated\\PSNR}}  \\
\midrule
\multirow{2}{*}{Matrix} &  ReSpeaker & 20.51 & 28.08 \\
&PlugUSB & 26.48 & 28.56  \\\\
\multirow{2}{*}{ReSpeaker} & Matrix & 20.51 & 24.15  \\
&PlugUSB & 21.65 & 28.23 \\\\
\multirow{2}{*}{PlugUSB} &Matrix  & 26.48 &  28.24 \\
&ReSpeaker & 21.65 & 24.21 \\
\bottomrule
\end{tabular}
\end{small}
\caption{Comparison of PSNR between the spectrograms coming from two different microphones before and after translation.}
\label{eg-table}
\vspace{-0.4cm}
\end{table}
{\subsecspace}
\subsection{Accuracy gains using \system{}}
{\subsecspace}
\label{subsec.gantraining}
\noindent
In this section, we evaluate the accuracy gains for audio models by incorporating \system{} in the inference pipeline. 

\begin{figure*}[t]
\captionsetup{font=small}
\begin{subfigure}[b]{0.3\linewidth}
\includegraphics[width=\linewidth]{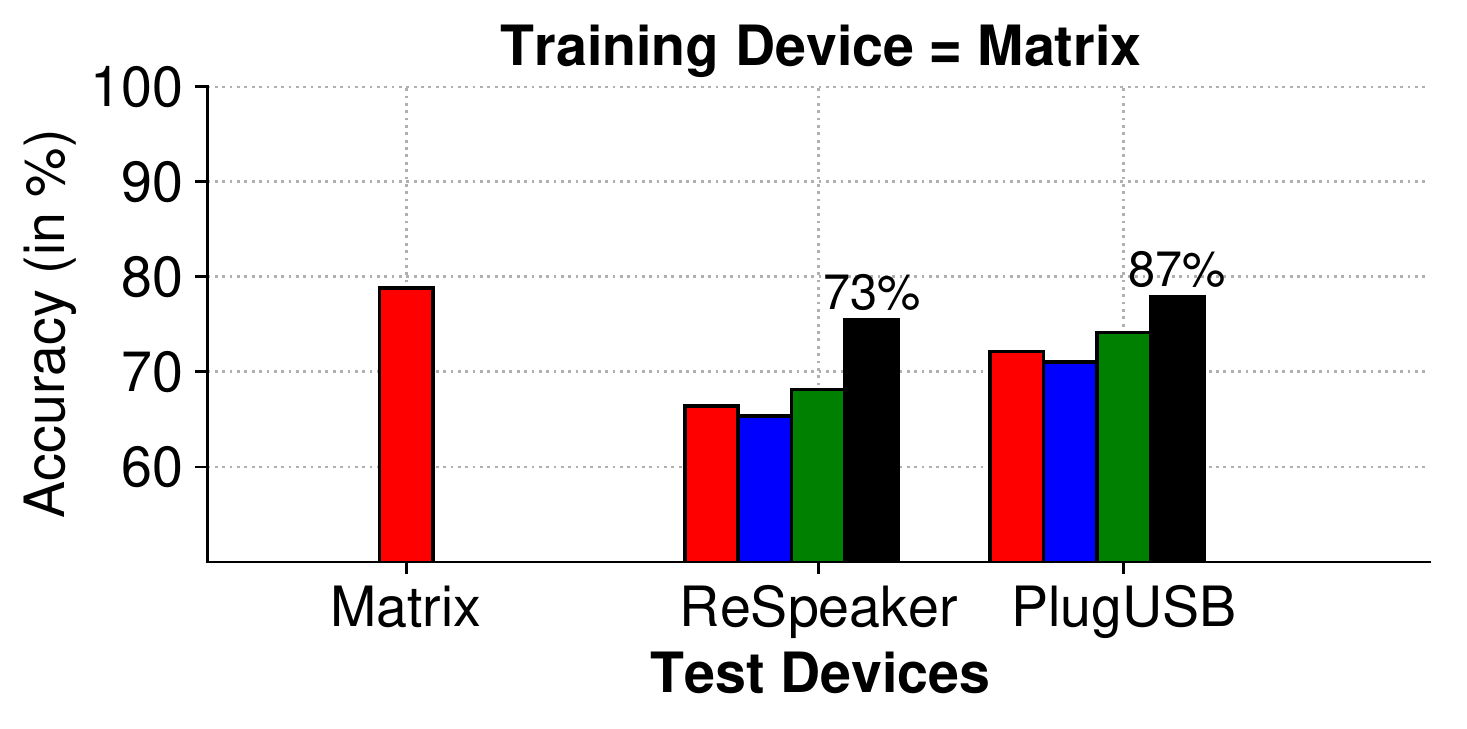}
\label{matrix-hotkey}
\vspace{-0.4cm}
\caption{}
\end{subfigure}
\begin{subfigure}[b]{0.3\linewidth}
\includegraphics[width=\linewidth]{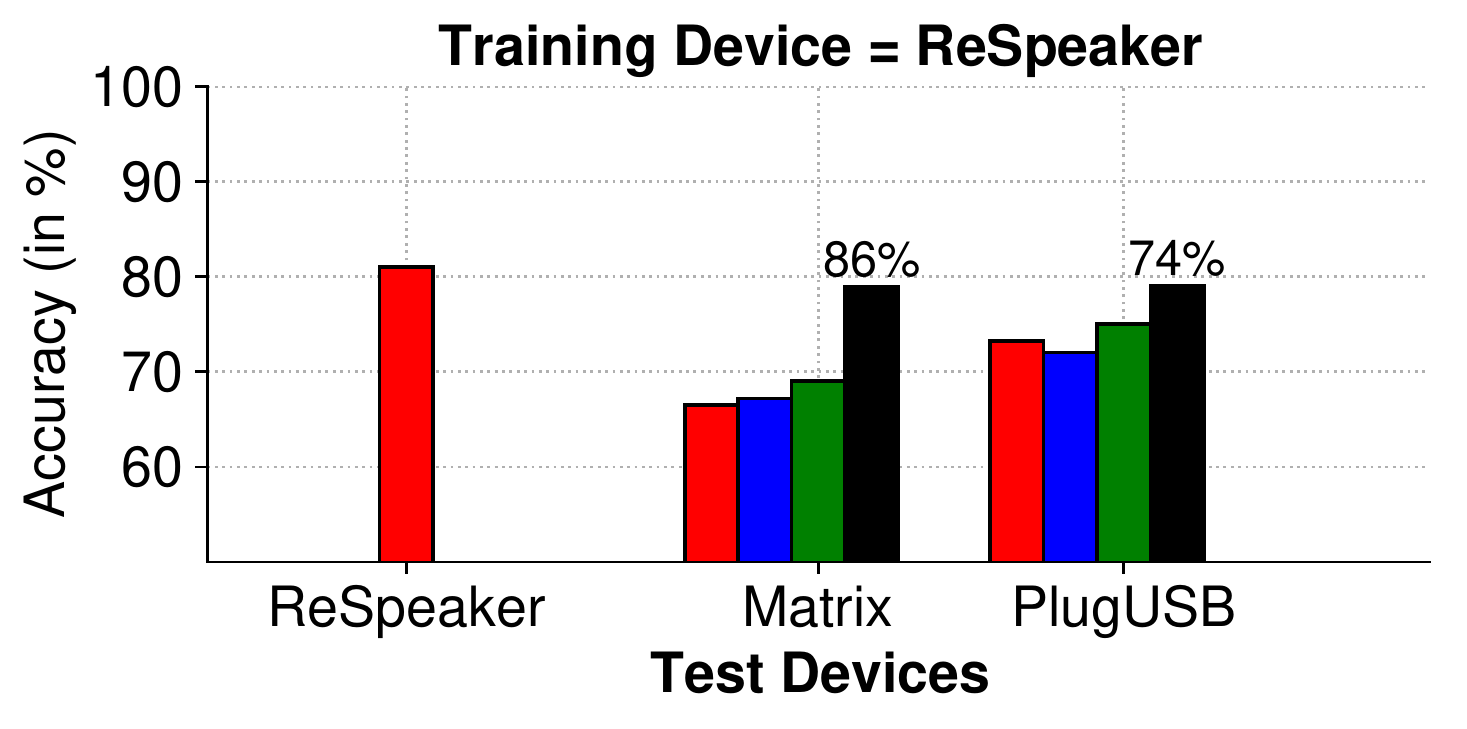}
\label{rs-hotkey}
\vspace{-0.4cm}
\caption{}
\end{subfigure}
\begin{subfigure}[b]{0.3\linewidth}
\includegraphics[width=\linewidth]{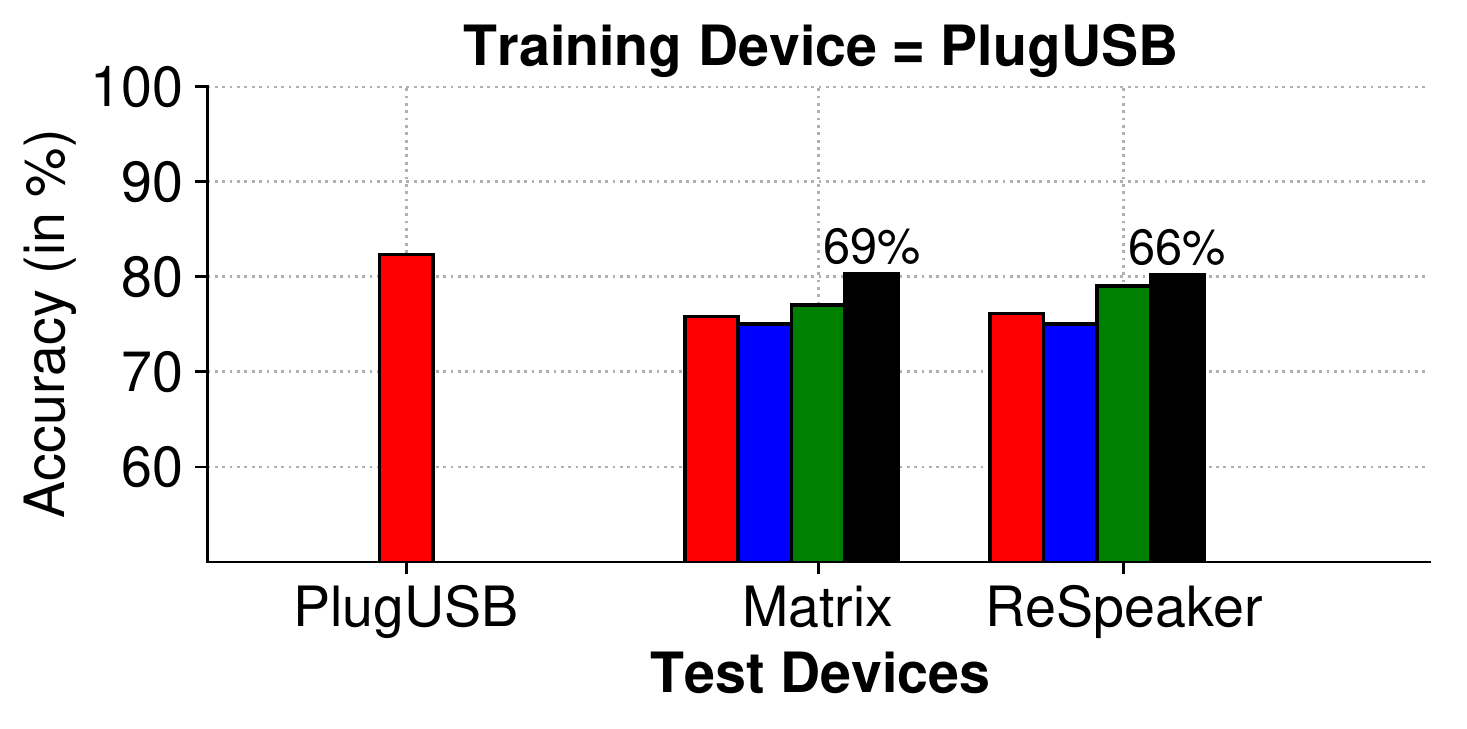}
\label{usb-hotkey}
\vspace{-0.4cm}
\caption{}
\end{subfigure}

\begin{subfigure}[t]{0.3\linewidth}
\includegraphics[width=\linewidth]{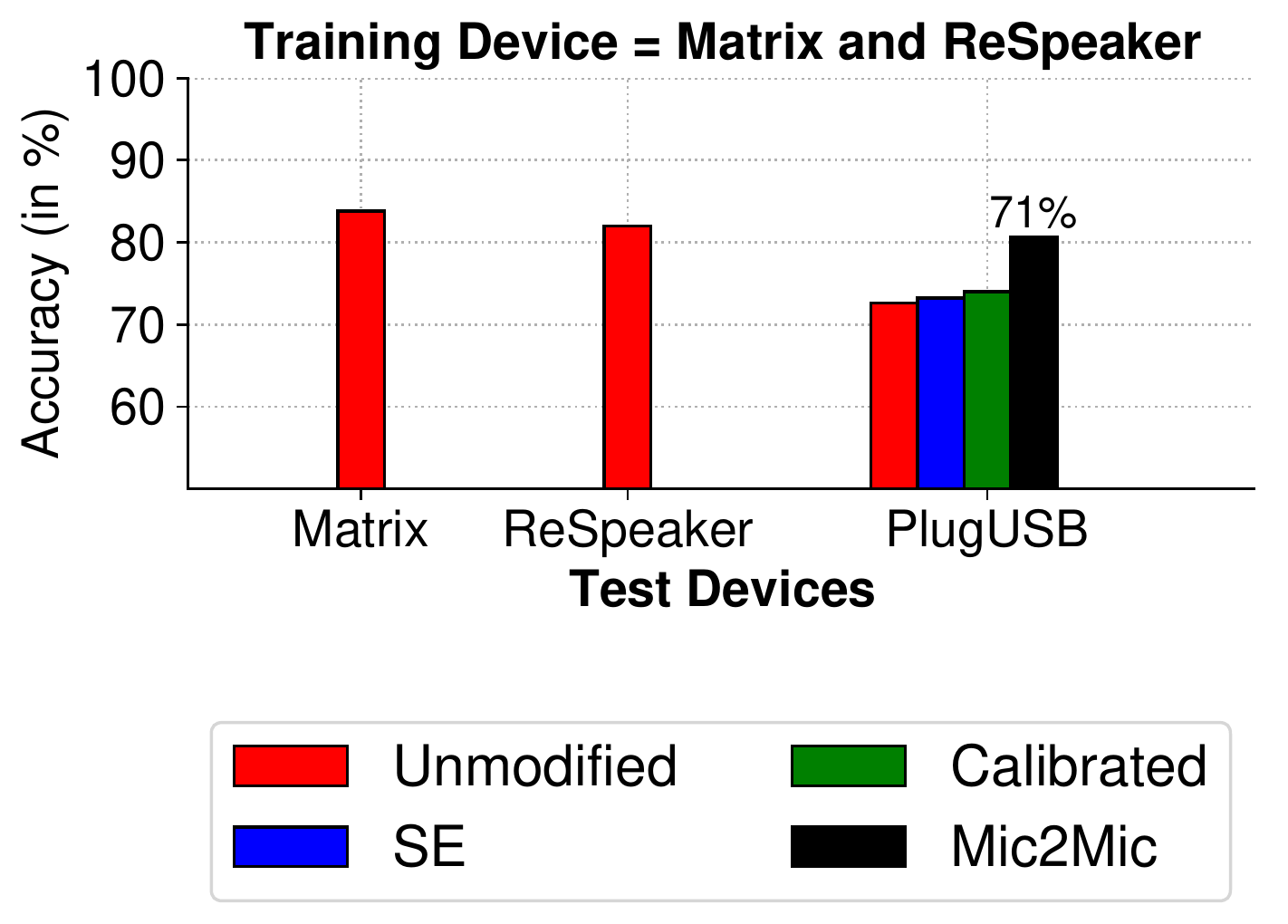}
\label{matrix-hotkey-multiple}
\vspace{-0.4cm}
\caption{}
\end{subfigure}
\begin{subfigure}[t]{0.3\linewidth}
\includegraphics[width=\linewidth]{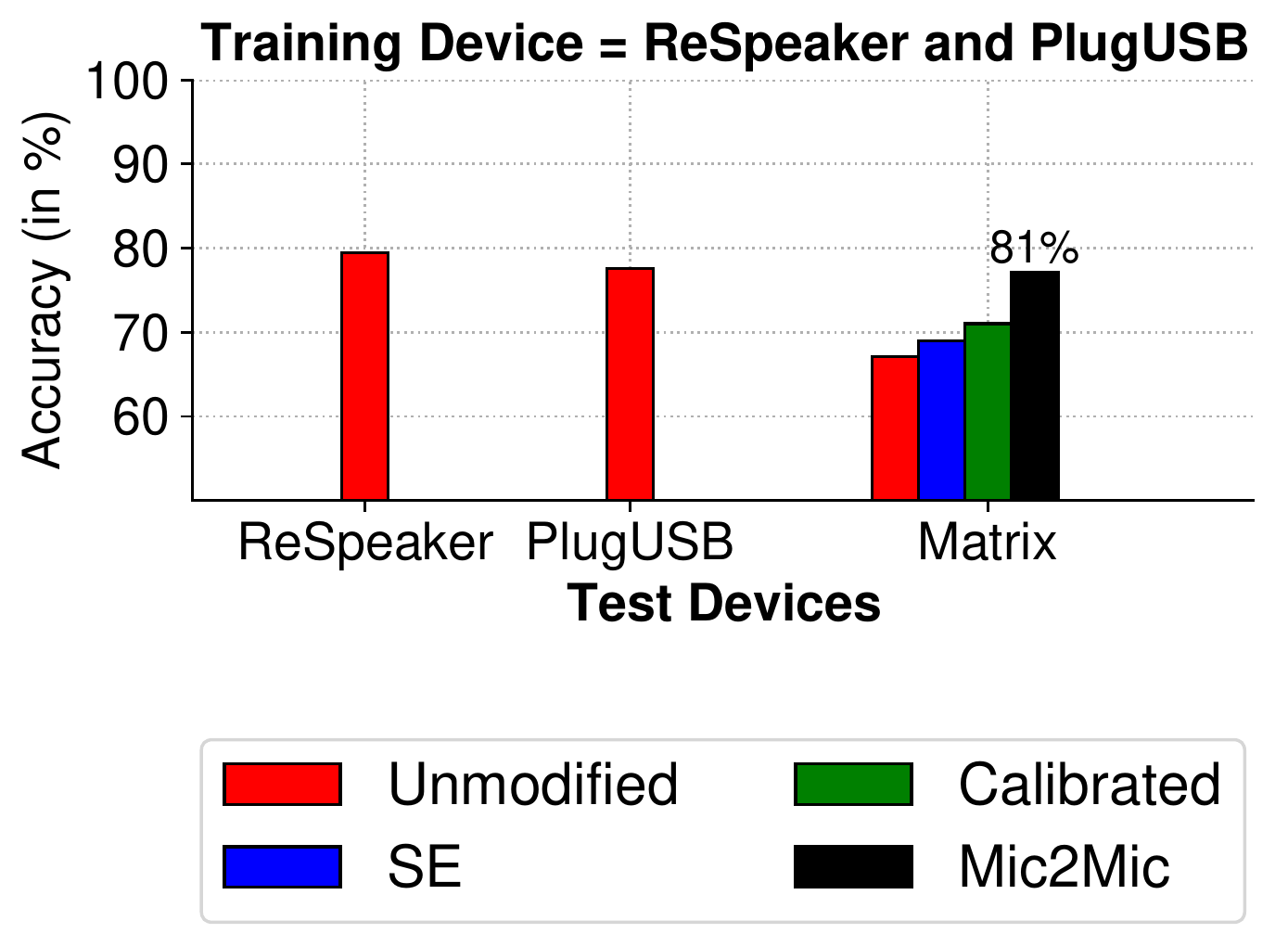}
\label{rs-hotkey-multiple}
\vspace{-0.4cm}
\caption{}
\end{subfigure}
\begin{subfigure}[t]{0.3\linewidth}
\includegraphics[width=\linewidth]{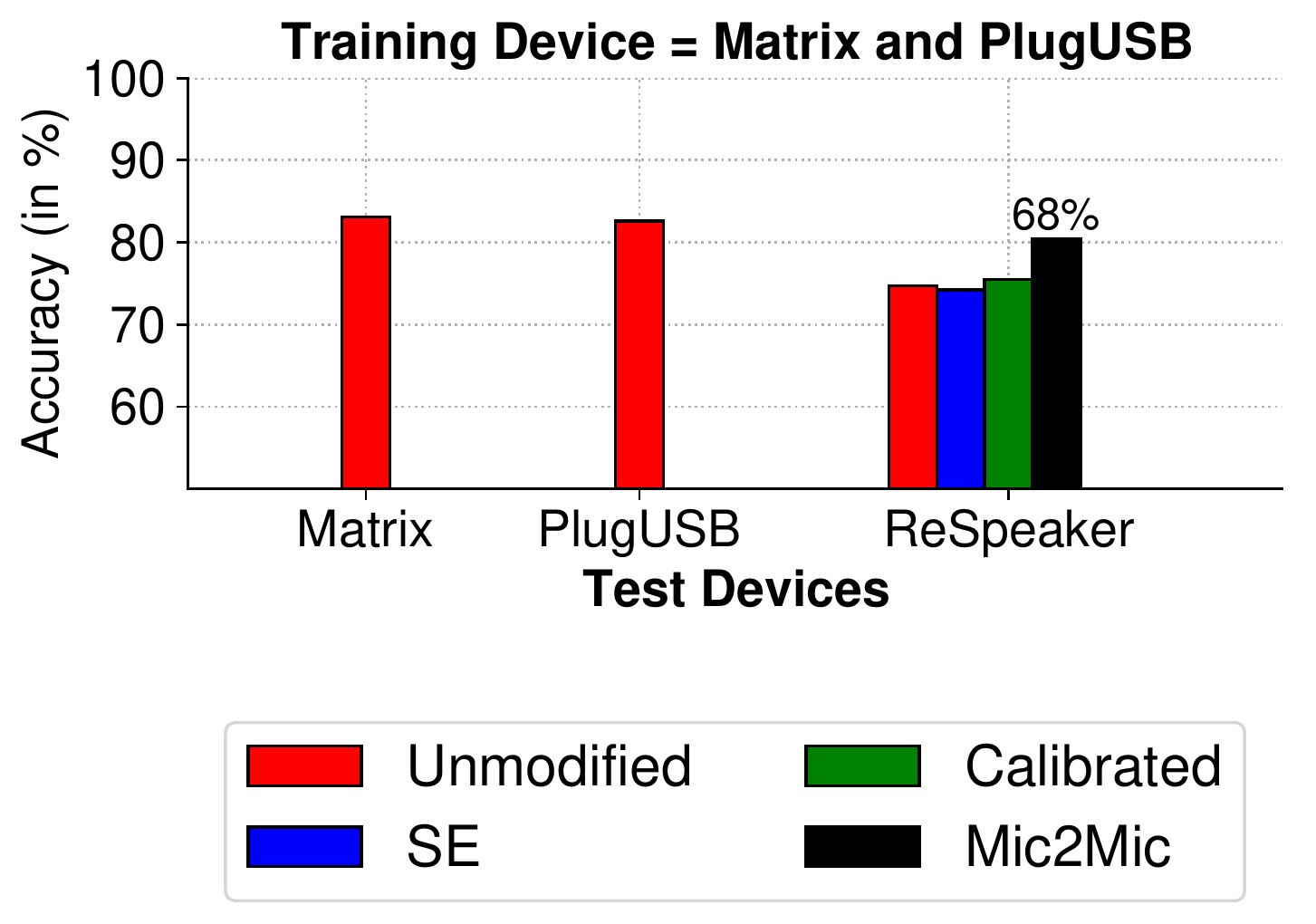}
\label{usb-hotkey-multiple}
\vspace{-0.4cm}
\caption{}
\end{subfigure}
\vspace{-0.4cm}
\caption{Accuracy of the Keyword Detection model under different scenarios of microphone variability. The numbers on the bars denote the percentage of lost accuracy recovered using \system{}.}
\label{hotkey}
\vspace{-0.4cm}
\end{figure*}

{\parjump}

\noindent
\textbf{Experiment Setup.} %
We present a series of experiments wherein the microphones used to train and test the audio models are different. We compare the performance of Mic2Mic against a number of baseline approaches in improving the robustness of audio models to such microphone variability.  

{\parjump}

\noindent
\textbf{Inference Pipelines.} %
When an audio model is deployed on a new microphone, we evaluate its performance in four scenarios:

 \squishlist {
 \item{\emph{Unmodified}: In this pipeline, the test data from the new microphone is directly passed to the audio model without applying any pre-processing technique.}

\item {\emph{Speech Enhancement (SE)}: Before feeding the test data to the audio model, we first enhance its speech quality by applying a statistical SE technique known as Minimum Mean-Square Error Short-Time Spectral Amplitude (MMSE-STSA) estimation~\cite{ephraim1984speech}.} 

\item{\emph{Calibrated}: An alternative approach to address microphone variability is to first measure the differences between training and test microphones in a controlled experimental setup and then negate those differences during the inference stage. While this approach of microphone calibration in controlled environments is not practical and scalable, we use it as a baseline to show how \system{} compares with it in terms of performance. To compute a calibration offset between a pair of microphones, we play a frequency sweep signal $X$ on a reference speaker and record it on our target microphones in a non-reflective anechoic chamber. 

The power spectral density $R_i$ of the microphone outputs can be expressed as follows, where $H_i$ is the transfer function of the $i^{th}$ microphone and $R_x$ is the PSD of the original frequency sweep signal. 

\vspace{-0.3cm}
\begin{equation}
    \begin{aligned}
	R_i(f) = {|H_i(f)|}^2 . R_x(f) 
    \end{aligned}
\label{eq:mic}
\end{equation}
\vspace{-0.35cm}

Thereafter, a calibration offset $\Gamma_{i, j}$ between two microphones $i$ and $j$ representing the difference in their frequency responses can be calculated as follows: 

\vspace{-0.3cm}
\begin{equation}
    \begin{aligned}
\Gamma_{i, j} = \frac{H_i}{ H_j} =  \sqrt{\frac{R_i(f)}{R_j(f)}}
    \end{aligned}
\label{eq:phi}
\end{equation}
\vspace{-0.35cm}

Finally, the pre-computed calibration offset $\Gamma$ is applied on the data from the test microphone to compensate for the differences in frequency response between the training and test mics.

 \item{\emph{\system{}}}: This pipeline follows our proposed approach as shown in Figure~\ref{diag:inference} wherein the data from the test microphone is first translated using \system{} and then passed to the task-specific audio model for inferences. We use 15 minutes of unpaired data from the training and test microphones to learn a translation model. Later in \S~\ref{subsec.data_scale}, we provide further results on how changing the amount of training data impact the performance of Mic2Mic.}
}
 \squishend

\begin{figure*}[t]
\captionsetup{font=small}
\begin{subfigure}[b]{0.3\linewidth}
\includegraphics[width=\linewidth]{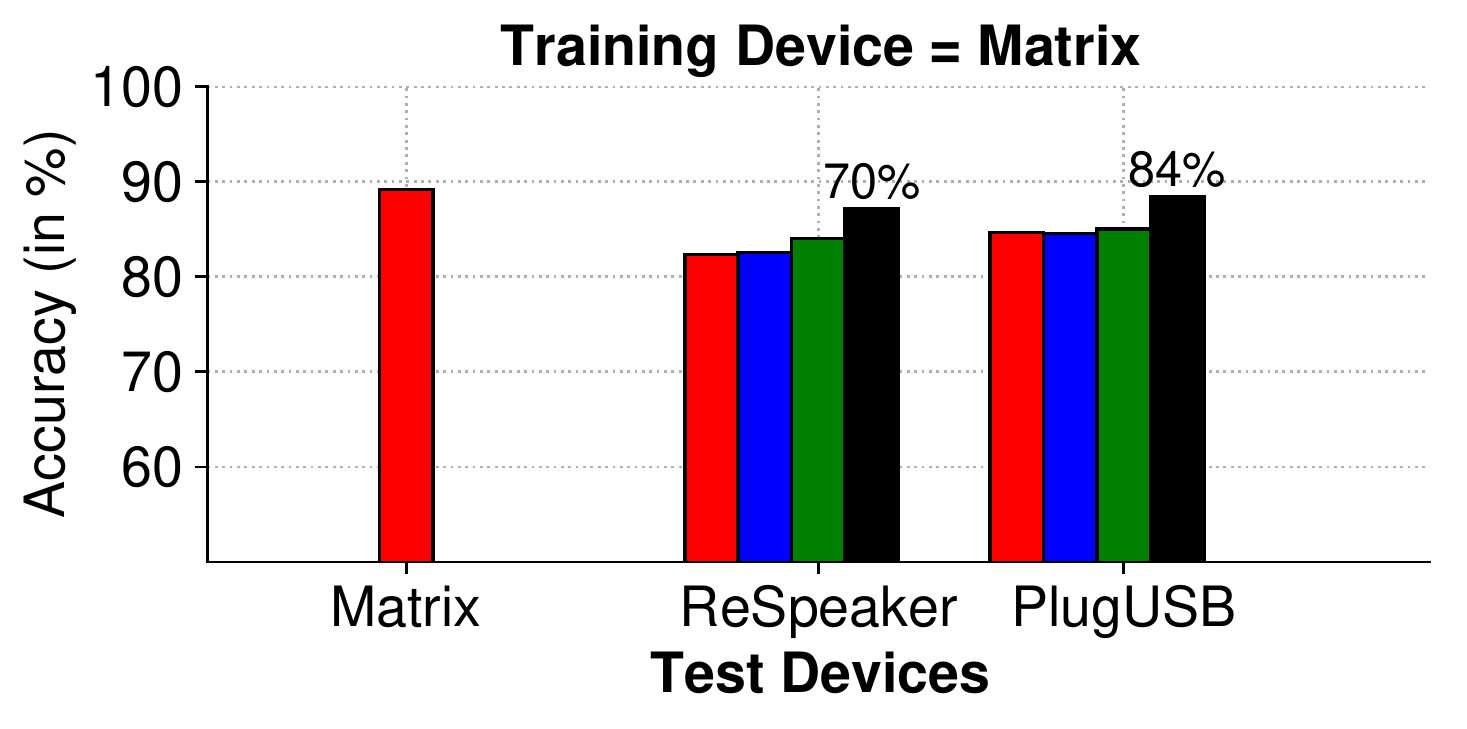}
\label{matrix-emotion}
\vspace{-0.4cm}
\caption{}
\end{subfigure}
\begin{subfigure}[b]{0.3\linewidth}
\includegraphics[width=\linewidth]{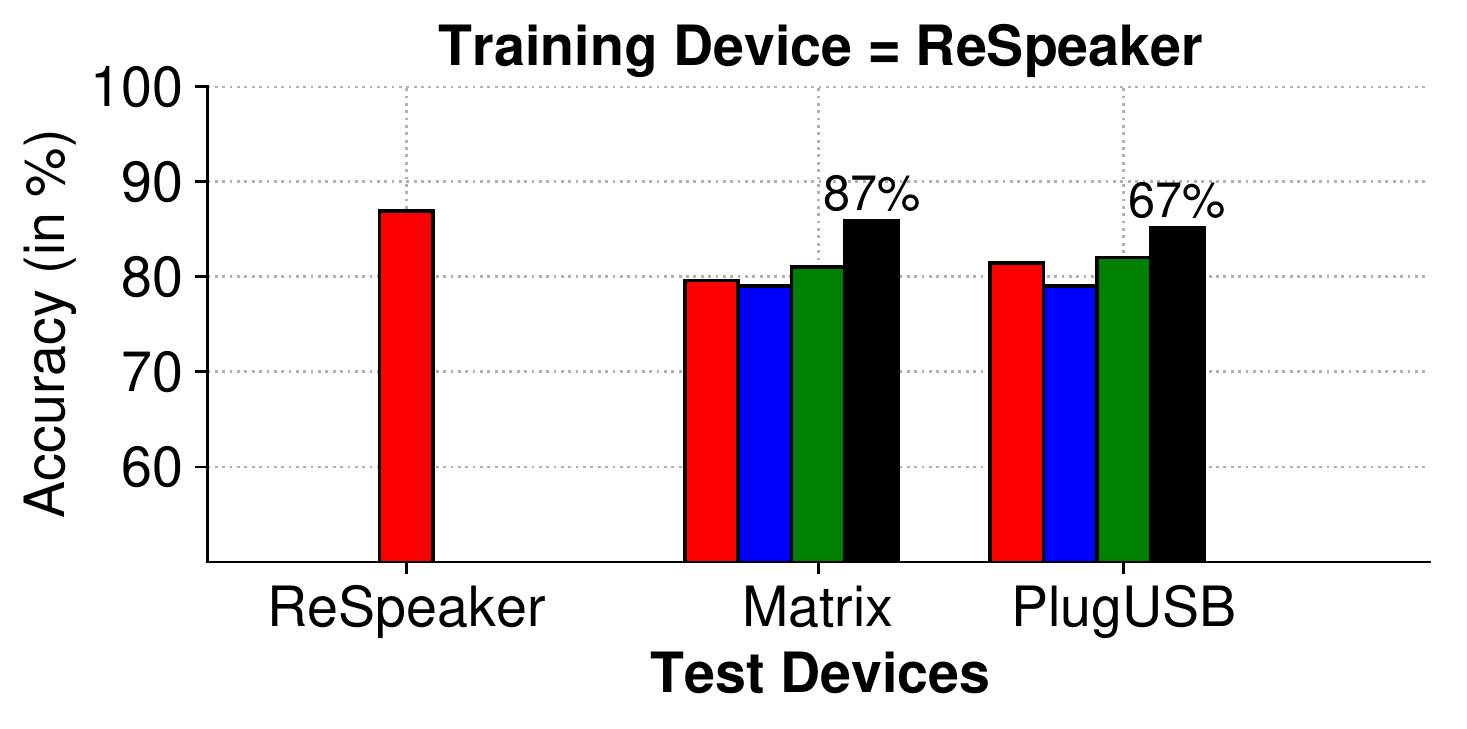}
\label{rs-emotion}
\vspace{-0.4cm}
\caption{}
\end{subfigure}
\begin{subfigure}[b]{0.3\linewidth}
\includegraphics[width=\linewidth]{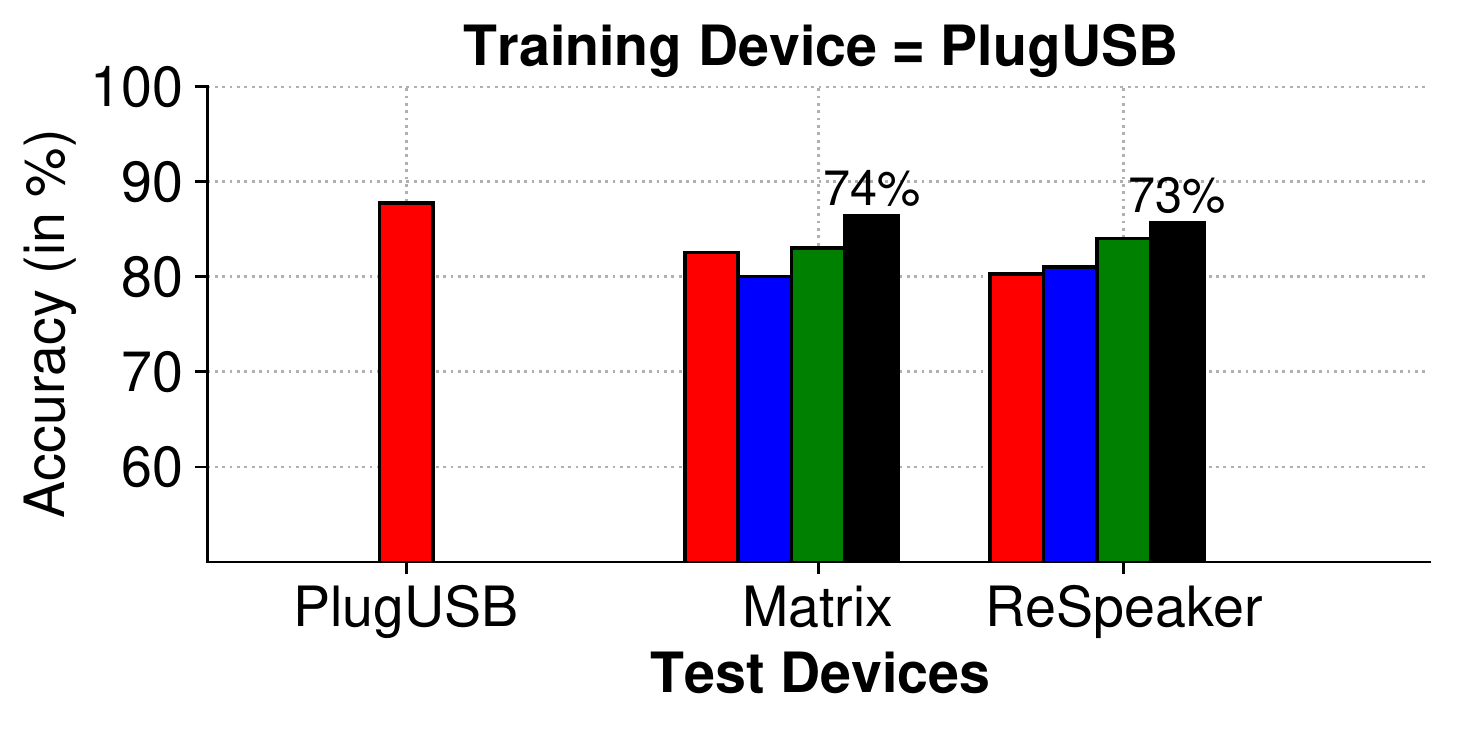}
\label{usb-emotion}
\vspace{-0.4cm}
\caption{}
\end{subfigure}

\begin{subfigure}[b]{0.3\linewidth}
\includegraphics[width=\linewidth]{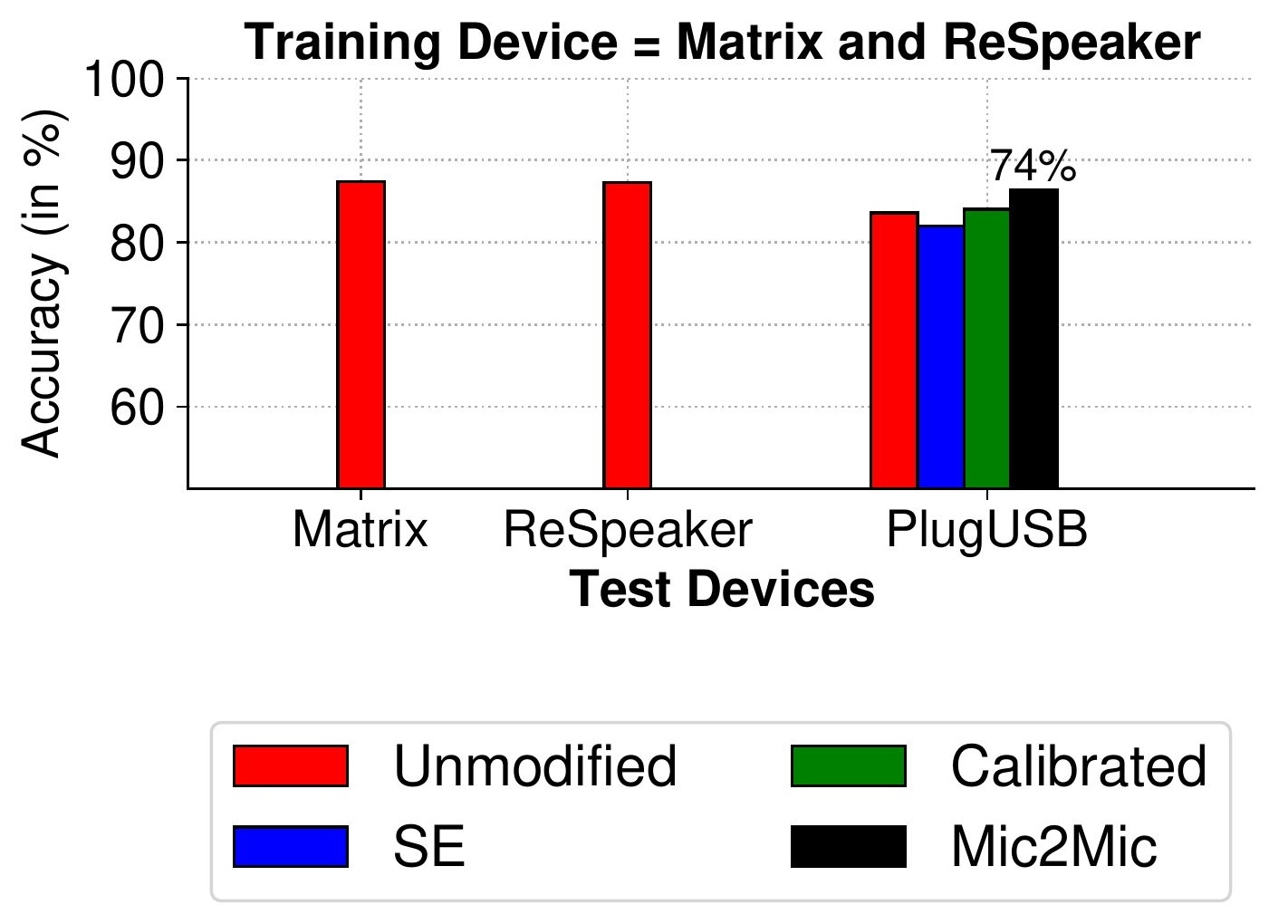}
\label{matrix-emotion-multiple}
\vspace{-0.4cm}
\caption{}
\end{subfigure}
\begin{subfigure}[b]{0.3\linewidth}
\includegraphics[width=\linewidth]{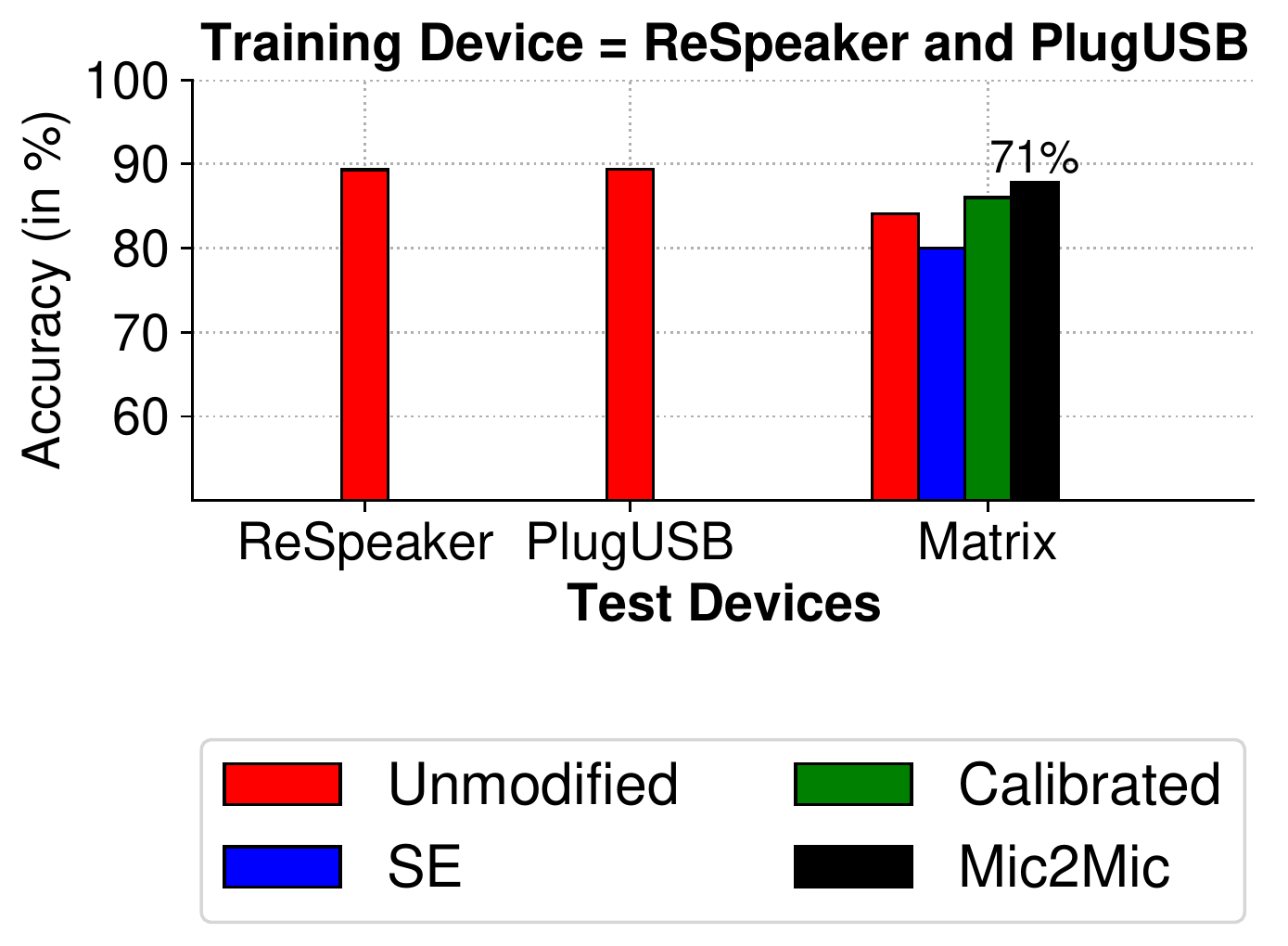}
\label{rs-emotion-multiple}
\vspace{-0.4cm}
\caption{}
\end{subfigure}
\begin{subfigure}[b]{0.3\linewidth}
\includegraphics[width=\linewidth]{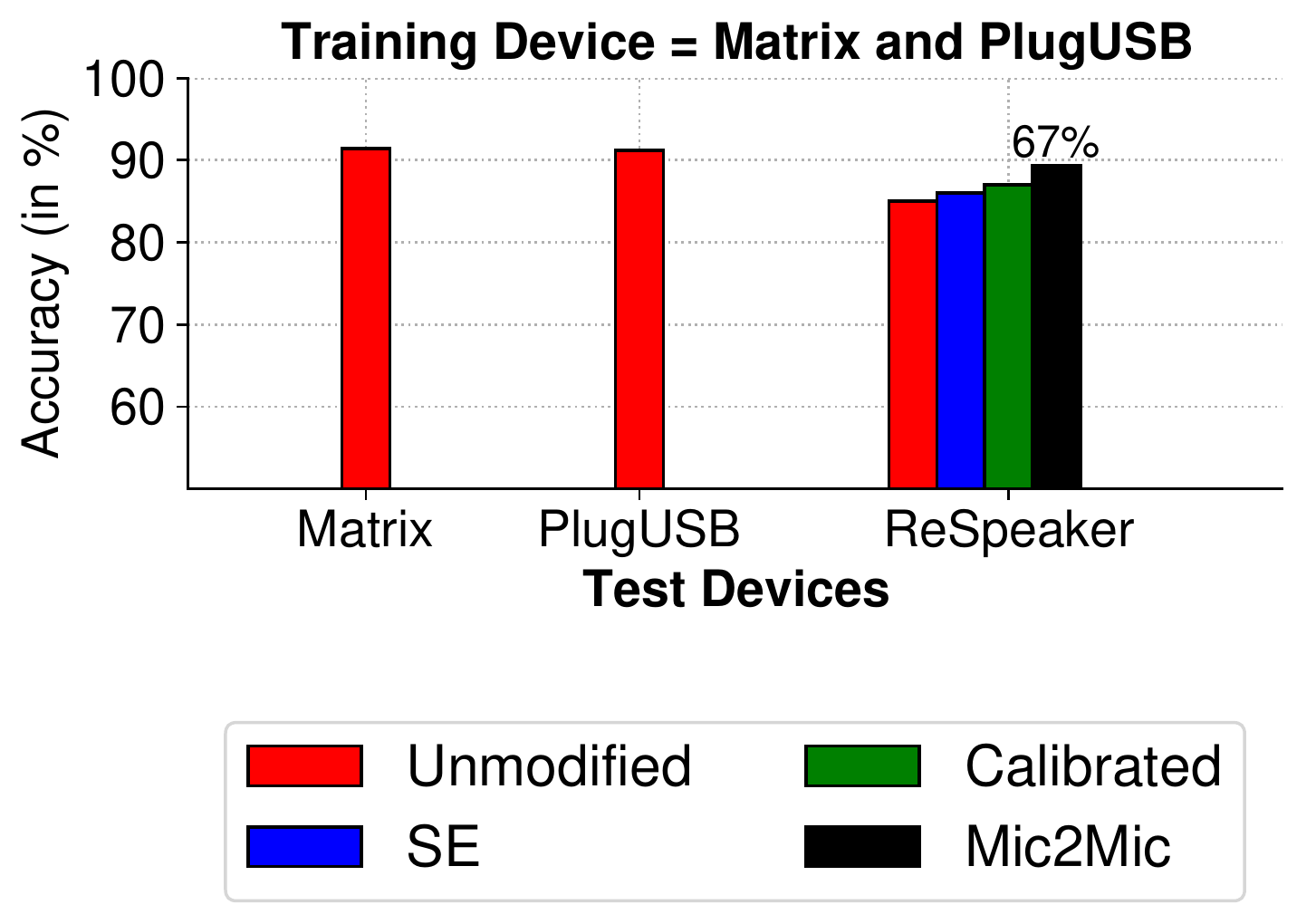}
\label{usb-emotion-multiple}
\vspace{-0.4cm}
\caption{}
\end{subfigure}
\vspace{-0.4cm}
\caption{Accuracy of the Emotion Detection model under different scenarios of microphone variability. The numbers on the bars denote the percentage of lost accuracy recovered using \system{}.}
\label{emotion}
\vspace{-0.4cm}
\end{figure*}

{\parjump}

\noindent
\textbf{Results.} %
In Figure~\ref{hotkey}, we present the mean accuracy scores on test microphones when the Keyword Spotting model is trained on the \emph{SC-12} dataset collected from different training microphones. As expected, we observe the upper bound inference accuracy when the model is tested on the same device on which it was trained, as there is no accuracy lost due to microphone variability. However when there is a mismatch between the training and test microphones, a significant accuracy loss is observed. For example, as shown in Figure~\ref{hotkey}a, when the model trained on Matrix Voice is deployed on ReSpeaker and PlugUSB microphones, there is an absolute accuracy drop of 12.4\% and 6.7\% respectively when compared with the accuracy upper bound (78.8\%). By incorporating \system{} in the inference pipeline, we are able to recover a significant portion of this accuracy loss (73\% and 87\%) respectively. In doing so, \system{} outperforms the two baseline approaches related to speech enhancement and microphone calibration. The superior performance of \system{} can also be observed in other training and test device combinations as shown in Figures~\ref{hotkey}(a-c). 

Similarly in Figure~\ref{emotion}, for the emotion detection task, we observe that accuracy drops between 4\%-9\% due to microphone variability and \system{} inference pipeline is able to recover above 80\% of the lost accuracy in some scenarios. 

We note that although Mic2Mic is able to recover a significant percentage of the lost accuracy (up to 87\%) due to microphone variability, it underperforms the ideal scenario in which training data from the target microphone in available. For example, in the Matrix to PlugUSB translation, \system{} improves the Keyword Detection accuracy from 72.1\% to 77.9\%, however it falls short of reaching the best-case accuracy of 82.3\% which could obtained with supervised training on PlugUSB. Future work can look at semi-supervised approaches where Mic2Mic is combined with a small amount of labeled data from the target microphone as a way to further improve its performance. 

{\parjump}
\noindent
\textbf{Training with multiple microphones}. While the above experiments use data from just one microphone for training the audio models, in practice developers will have access to data from diverse microphones. We now evaluate if accuracy drops still persist when an audio model is trained with multiple microphones and tested on a new microphone. 

In Figures~\ref{hotkey}(d-f) and ~\ref{emotion}(d-f), we present the mean accuracy scores when the audio models are trained on data from multiple training microphones and deployed on an unseen test microphone. We again observe an accuracy drop due to training and test microphone mismatch (e.g., the accuracy drops by ~10\% when Keyword Detection model trained on Matrix and PlugUSB is deployed on ReSpeaker). Mic2Mic inference pipeline is able to recover nearly 70\% of this lost accuracy in all scenarios.

\begin{table}[t]
\captionsetup{font=small}
\centering
\begin{small}
\begin{tabular}{lccccc}
\toprule
{\parbox{1cm}{\centering Training\\Microphones}} & {\parbox{1cm}{\centering Test\\Microphones}} & Unmodified & DA & GAN-paired & Mic2Mic   \\
\midrule
\multirow{3}{*}{Matrix} & Matrix & 78.8 & NA & NA & NA \\
&ReSpeaker & 66.40 & 70.84 & \textbf{75.64} & 75.51 \\
&PlugUSB & 72.1& 74.5 & 77.48 & \textbf{77.9} \\\\
\multirow{3}{*}{ReSpeaker} & ReSpeaker & 81 & NA & NA & NA\\
&Matrix & 66.5 & 70.47 & 76.56 & \textbf{78.9} \\
&PlugUSB & 73.22 & 76.37 & \textbf{80.4} & 79 \\\\
\multirow{3}{*}{PlugUSB} & PlugUSB & 82.3 &NA & NA & NA\\
&Matrix  & 75.8 &  77.8& 79.89 & \textbf{80.3} \\
&ReSpeaker & 76.1 & 77.45 & 79.38 & \textbf{80.2} \\
\bottomrule
\end{tabular}
\end{small}
\caption{Comparison of Mic2Mic against paired training approaches.}
\label{da-table}
\vspace{-0.8cm}
\end{table}

{\parjump}
\noindent
\textbf{Comparison with Paired Approaches}. We now compare the performance of \system{} against techniques which rely on the availability of paired data from source and target microphones. As explained earlier, the need for paired data is a strong assumption which is not practical in real-world scenarios, however it may be possible to do paired data collection in a lab setting, particularly if the number of deployment devices are small and the sensing task does not require large-scale data. 

Our first paired baseline is a recently proposed approach~\cite{ipsn18_mathur} which uses data augmentation (DA) training as a way of regularization to improve the accuracy of audio models trained for smartphones and deployed on smartwatches. To do so, it computes the relative transfer function between microphones, and use it for creating an augmented training dataset upon which the target audio model is retrained. Further, we benchmark \system{} against a GAN-based training approach where the GAN is trained to learn a translation function using paired microphone data. To this end, we use the `paired' version of the Speech Commands dataset collected in our experiments. This approach differs from \system{} in two ways: a) for training \system{}'s CycleGAN, we ensure that no paired data is passed to the training algorithm whereas in the GAN-paired baseline, we enforce that $p(x_A)$ and $p(x_B)$ are paired. b) we do not use the cycle-consistency loss in the training process. 

We implement these paired baselines in the context of the Keyword Spotting task and present the results in Table~\ref{da-table}. We observe that although the data augmentation (DA) approach trains a model that beats the baseline (unmodified inference pipeline), Mic2Mic is able to outperform this approach for all combinations of microphones. Further, our results show that Mic2Mic and GAN-paired approach provide similar performance improvements in most cases, however we observe that the GAN-paired technique has ~2x faster training convergence than the unpaired Mic2Mic approach.

{\subsecspace}
\subsection{How much data is needed to train \system{}?}
{\subsecspace}
\label{subsec.data_scale}
\noindent

In this section, we present results on \system{}'s performance as the amount of training data is varied. As discussed, training \system{} only requires unlabeled and unpaired data from the source and target microphones. While collecting such data is indeed cheap, it will be ideal if the \system{} can be trained with minimal amount of data.

\begin{figure}[htb]
\captionsetup{font=small}
\begin{subfigure}[b]{0.49\linewidth}
\includegraphics[width=\linewidth]{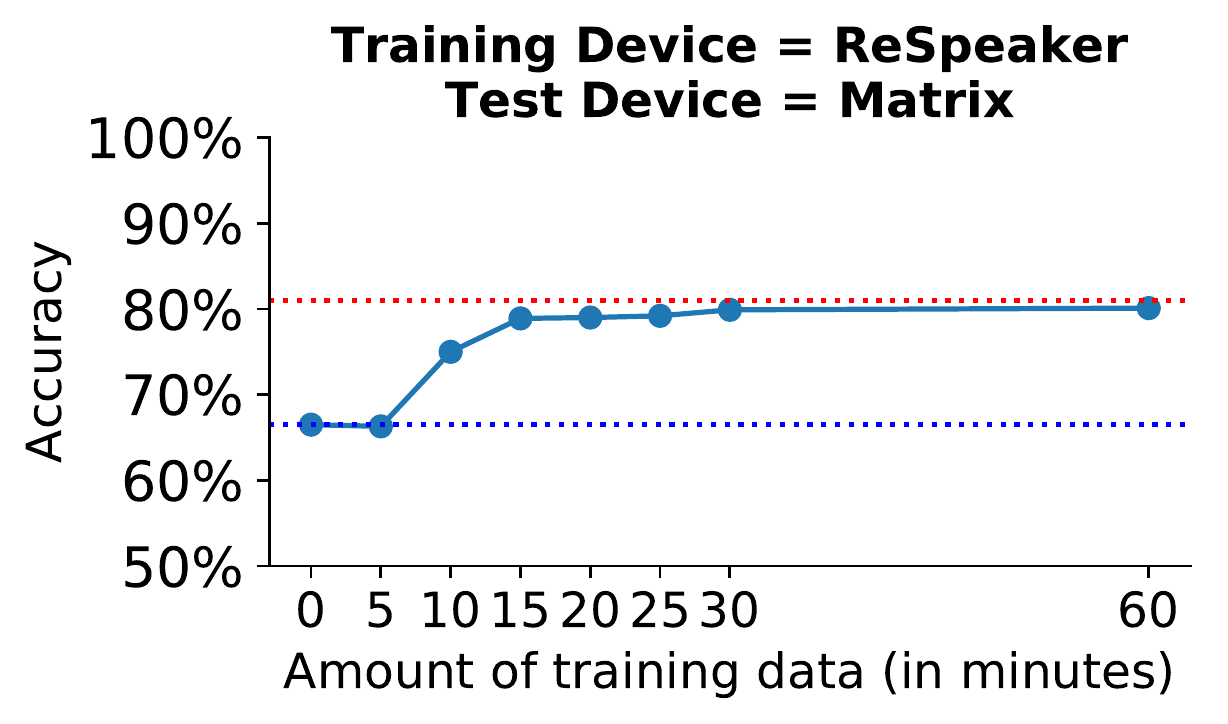}
\label{amount1}
\vspace{-0.4cm}
\caption{}
\end{subfigure}
\begin{subfigure}[b]{0.49\linewidth}
\includegraphics[width=\linewidth]{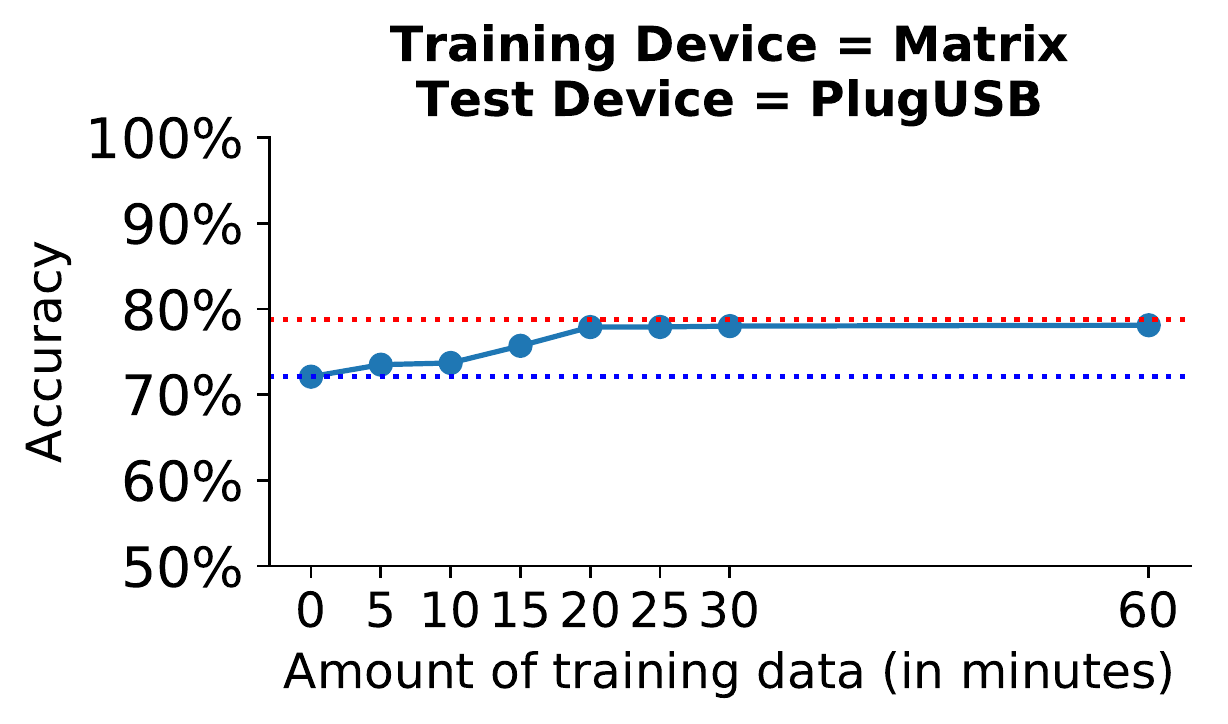}
\label{amount2}
\vspace{-0.4cm}
\caption{}
\end{subfigure}
\vspace{-0.4cm}
\caption{Inference accuracy of the Keyword Detection model as the amount of data used to train the \system{} translation model is varied. The inference accuracy plateaus after the CycleGAN is trained with around 18 minutes of data.}
\label{data-amount}
\vspace{-0.4cm}
\end{figure}

{\parjump}

\noindent
\textbf{Experiment Setup.} %
We systematically vary the amount of unpaired data available to train \system{}'s translation model. We gradually increase the unpaired data from the training and test microphones from 0 minutes to 60 minutes. Once the CycleGAN is trained in each data configuration, we incorporate it in the inference pipeline of the Keyword Spotting model and evaluate the accuracy gains. 


{\parjump}

\noindent
\textbf{Results.} %
Figure~\ref{data-amount} illustrates the findings of this experiment for two different microphone combinations. The dotted-red line shows the upper-bound test accuracy when the Keyword Spotting model is tested on the same microphone on which it was trained. The dotted blue line shows the baseline accuracy when the model is deployed on a test microphone without incorporating Mic2Mic in the inference pipeline.  

We observe that as more \emph{unpaired} training data is supplied, the Mic2Mic translation model becomes better and in turn, the inference performance of the Keyword Spotting model increases. The inference performance plateaus around 15 minutes in Figure~\ref{data-amount}(a) and around 18 minutes in Figure~\ref{data-amount}(b), suggesting that Mic2Mic is able to learn a good translation model with less than 20 minutes of unpaired data in these scenarios. 


{\subsecspace}
\subsection{System Evaluation}
{\subsecspace}
\label{subsec.system}
\noindent

We now evaluate the runtime performance of \system{}'s translation model on a number of mobile and embedded platforms. For this experiment, we executed the translation model only on the CPU and did not use other on-board processors such as the GPU or the DSP. Table~\ref{tbl.system} shows the latency of translating a 5 second audio on different processors. We observe that it takes around less than 150ms to translate a 5 second audio segment on newer mobile processors, however on older processors such as Snapdragon 400, it may take up to 750ms. In future, we will investigate avenues to optimize the runtime of the convolutional operations~\cite{bhattacharya2016sparsification} in our translation model in order to further reduce its execution latency on embedded devices. Further, as modern processors also have on-board GPU, we can leverage it to reduce the translation time.

\begin{table}
\centering
\singlespacing
\tabcolsep=0.11cm
\begin{tabular}{|c|c|c|} \hline
\thead{Processor} & \thead{Release Year} & \thead{Latency} \\ \hline
Snapdragon 835 & 2017 & 112ms \\\hline
Snapdragon 820 & 2016 & 135ms \\\hline
Snapdragon 410 & 2016 & 270ms \\\hline
Snapdragon 400 & 2014 & 750ms \\\hline

\end{tabular}
\caption{Execution latency of Mic2Mic in translating a 5 second audio segment. }
\vspace{-0.4cm}
\label{tbl.system}
\end{table}

{\secspace}
\section{Discussion and Limitations}

\label{sec:dis}
{\secspace}
\noindent
In this section, we discuss the limitations of our work, and outline the avenues for future work on this topic.  

{\parjump}

\noindent
\textbf{Scalability of Pairwise Translations.} A limitation of our current implementation is that we learn pairwise translations between training and test microphones, as such in order to scale this solution to a large number of devices, multiple such models need to be learned. As a future work, we are investigating the development of a common microphone translation model that can be applied to any given pair of microphones. Recently, a similar solution in the image-to-image translation domain has been proposed, namely StarGAN~\cite{choi2017stargan} which uses a n-dimensional one-hot vector to encode the input and output image labels in the model architecture. 

{\parjump}
\noindent
\textbf{Deployment Overhead.} \system{} adds an extra step of data translation in the inference pipeline of audio models. In future, we will explore avenues for optimizing the translation model to reduce its latency and power consumption on embedded devices. As our translation model consists of a number of convolution layers, we plan to use techniques proposed for optimizing CNNs on embedded devices~\cite{bhattacharya2016sparsification}.

{\parjump}

\noindent
\textbf{Training with Diverse Microphones.} Our experiments showed that microphone diversity in the training set, albeit useful, is not sufficient to counter the challenge of domain shift. When the deployment microphone is different from the \emph{diverse} training microphones, approaches such as Mic2Mic are still useful. However, in the rare case when data from all deployment microphones are available while training, we expect that domain adaptation techniques such as Mic2Mic are not needed.  

{\parjump}
\noindent
\textbf{Controlled Study Setup.} Audio model robustness in real-world is a challenging, multifaceted problem. We study the challenge of microphone variability, therefore we controlled for other potential variabilities such as ambient noise, speaker accents etc. However when \system{} is deployed in the real-world, it will have to encounter other such forms of noise while learning a translation model. In future work, we will study the generalizability of Mic2Mic when the unpaired training data also contains other forms of noise besides microphone variability.

%
%

{\secspace}
\section{Related Work}
\label{sec:related}
{\secspace}
\noindent
We now review prior works on sensor variability in embedded devices, audio enhancement approaches and the use of GANs in audio applications. 

{\parjump}

\noindent
\textbf{Sensor Heterogeneity in Embedded Devices.} %
A number of past works have found that the accuracy of sensor classification models degrades in the real-world, owing to the variations in sensor devices~\cite{amft2010need, blunck2013heterogeneity}. Blunck et al. \cite{blunck2013heterogeneity} showed the impact of a GPS sensor variability on the data quality and the performance of inference models on smartphones. Stisen et al.~\cite{stisen2015smart} highlighted that even software factors such as CPU loads can cause a large variability in the accelerometer outputs of smartphones and smartwatches. In other works, ~\cite{ipsn18_mathur} proposed a data augmentation based training approach to counter device heterogeneities, while \cite{lu2010jigsaw} presented a number of user-driven and automatic methods to calibrate accelerometer responses from various target devices.

{\parjump}

\noindent
\textbf{Audio Enhancement for Speech Models.} %
In order to make audio models robust against real-world noise, a number of speech and audio enhancement techniques have been proposed. Classic techniques for speech enhancement include Wiener filtering~\cite{lim1978all}, subspace algorithms~\cite{ephraim1995signal}, and statistical methods such as STSA-MMSE~\cite{ephraim1984speech}. Recently, deep learning has been prominently use to tackle this problem with solutions ranging from simple feed-forward neural networks~\cite{lane2015deepear}, denoising auto-encoders~\cite{lu2013speech}, and RNNs with LSTM units~\cite{maas2012recurrent, weninger2015speech}. The common approach adopted by these works is to first add artificial noise to clean audios, and then feed the time-aligned clean and noisy audio pairs to a neural network, for it to learn a mapping function from noisy to clean audios. As such, these approaches are not suitable for our problem domain, because collecting perfectly time-aligned paired data from multiple microphones is very challenging. Our proposed solution of using CycleGANs can learn mappings between microphones using unpaired and unlabeled audio samples. 

{\parjump}

\noindent
\textbf{Audio and Speech applications of GANs.} %
In the last few years, GANs have been used extensively for generating images~\cite{liu2017auto,isola2017image}. Only recently, there has been an increased focus on using GANs for speech enhancement and speech generation. ~\cite{pascual2017segan} proposed an application of GANs for speech enhancement by operating them on raw audios. Michelsanti et al.~\cite{michelsanti2017conditional} adapted a conditional GAN designed for image-to-image translation for the task for speech enhancement and speaker verification. ~\cite{donahue2018synthesizing} introduced WaveGAN which can be used to synthesize diverse audios such as bird vocalizations, drums, and piano sounds. However, none of these solutions tackle the problem of microphone variability on embedded devices, which is the primary focus on our work.

{\secspace}
\section{Conclusion}
\label{sec:conclusion}
{\secspace}
In this work, we examined the impact of microphone variability on the robustness of audio-based computational models. After systematically exploring the runtime behavior of multiple audio models under different microphone environments, we propose a machine-learned system component, namely \emph{Mic2Mic} which can translate audio data between microphone domains. Our results show that Mic2Mic can recover between 66\% and 89\% of the accuracy drop caused by microphone variabilities and it consistently beats a number of signal processing and calibration baselines.  

\balance
\bibliographystyle{ACM-Reference-Format}
\bibliography{refs} 

\end{document}